\documentclass[12pt,a4paper]{article}
\usepackage[onehalfspacing]{setspace}
\usepackage{amsmath,mathtools,amssymb,amsthm}
\usepackage[authoryear]{natbib}
\usepackage{authblk}

\usepackage{algorithm}
\usepackage{algpseudocode}
\usepackage{caption,subcaption,tabularx,booktabs}
\usepackage[]{threeparttable}
\usepackage{rotating}

\DeclareMathOperator*{\argmax}{arg\,max}
\DeclareMathOperator*{\argmin}{arg\,min}
\newcommand\fnote[1]{\captionsetup{font=footnotesize}\caption*{#1}}

\usepackage{comment,color}
 \includecomment{comment}
 \specialcomment{comment}{\begingroup\color{red}}{\endgroup}  

\title{Scalable Bayesian estimation in the multinomial probit model\thanks{We would like to thank an editor, an associate editor, and an anonymous referee for very constructive comments. We would also like to thank Gael M. Martin, David T. Frazier, Richard Paap, and Michael S. Smith for helpful discussions. Rub\'en Loaiza-Maya is an associate investigator with the Australian Centre of Excellence for Mathematical and Statistical Frontiers.}}
\date{\small\today\vspace{-0.7cm}}
\author{Rub\'en Loaiza-Maya}
\author{Didier Nibbering\thanks{Correspondence to: Department of Econometrics \& Business Statistics, Monash University, Clayton VIC 3800, Australia, e-mail: \textsf{didier.nibbering@monash.edu}}}
\affil{\small Department of Econometrics and Business Statistics, Monash University\vspace{-1.5cm}}
\author{}

\begin{document}
\maketitle
\begin{abstract} 
\noindent \footnotesize
The multinomial probit model is a popular tool for analyzing choice behaviour as it allows for correlation between choice alternatives. Because current model specifications employ a full covariance matrix of the latent utilities for the choice alternatives, they are not scalable to a large number of choice alternatives. This paper proposes a factor structure on the covariance matrix, which makes the model scalable to large choice sets. The main challenge in estimating this structure is that the model parameters require identifying restrictions. We identify the parameters by a trace-restriction on the covariance matrix, which is imposed through a reparametrization of the factor structure. We specify interpretable prior distributions on the model parameters and develop an MCMC sampler for parameter estimation. The proposed approach significantly improves performance in large choice sets relative to existing multinomial probit specifications. Applications to purchase data show the economic importance of including a large number of choice alternatives in consumer choice analysis.
\end{abstract}

\thispagestyle{empty}
\clearpage
\setcounter{page}{1}

\section{Introduction}
The multinomial probit (MNP) is an important model for analysing choice behavior, because it allows the latent utilities of the choice alternatives to be correlated. These correlations capture general substitution patterns among choice alternatives, in contrast to the case with models that impose the independence of irrelevant alternatives property \citep{hausman1984specification}.

However, current specifications of the multinomial probit model are not scalable to discrete choice problems with a large number of choice alternatives, as the number of parameters in the covariance matrix of the latent utitilities grows quadratically in the number of choice alternatives \citep{burgette2013multiple}. This curse of dimensionality is exacerbated by the fact that, contrary to standard covariance matrix estimation settings, where multiple continuous variables are observed, all parameters in the covariance matrix have to be estimated from a single categorical variable. 

Standard dimension reduction techniques, such as factor analysis, cannot straightforwardly be applied to the covariance matrix of the latent utitilities. Since the scale of the latent utilities is not identified \citep{bunch1991estimability}, the identification of the model parameters requires a restriction on the covariance matrix. The main challenge is to reduce the number of parameters that characterize the covariance matrix, while imposing an identifying restriction.

This paper proposes a multinomial probit model specification that is scalable to modern choice data with many choice alternatives. Specifically, we employ a factor structure on the covariance matrix, where the number of parameters scales linearly, rather than quadratically, with the number of choice alternatives. The parameters are identified by imposing a trace restriction on the covariance matrix. To impose the trace restriction on the factor representation, we transform the covariance parameters to a spherical coordinate system of angles and a spherical radius. The radius is, by construction, equal to the square root of the trace of the covariance matrix. Therefore, the trace restriction is readily imposed by setting the squared radius equal to the number of choice alternatives. 

To conduct Bayesian estimation, prior densities on the angles in the reparameterization must be selected. We elicit the priors on the angles from well understood prior assumptions popularly used in the Bayesian factor analysis literature. The process of elicitation can be performed using a fast algorithm provided in this paper. The computation of the posterior distribution involves a Markov Chain Monte Carlo (MCMC) sampler with Gibbs sampling steps for the coefficients and latent utilities, and a Metropolis-Hastings step for the angle parameters. A numerical experiment confirms that the MCMC sampler succeeds in accurately estimating the model parameters, in similar computation time as existing MNP specifications. 

An application to real consumer choice data illustrates the empirical relevance of the scalable multinomial probit model. We construct consumer choice data with 50 alternatives as a modern counterpart of commonly used laundry detergent and margarine purchase data sets with only six alternatives. In these large choice sets, our approach produces better predictive performance than existing methods. The model is able to identify correlations across a large set of products. The results show that limiting the analysis to only a few products may, for instance, severely bias price elasticity estimates. The proposed model has similar performance to existing multinomial probit specifications when applied to the traditional laundry detergent and margarine choice data with six alternatives. 

This paper makes three important contributions to the multinomial probit literature. First, the proposed approach addresses the scalability of the multinomial probit model directly. \citet{piatek2017multinomial} specify a factor structure for the covariance matrix under the assumption that the factor loadings are known. This assumption might be unrealistic in many choice problems, especially when the number of alternatives is large. \citet{cripps2009parsimonious} propose a covariance selection prior that permits elements of the inverse of the covariance matrix to be zero. This approach allows for a sparse representation of the model, but does not reduce the number of parameters to be estimated. 

Second, this paper contributes to the literature on parameter identification in multinomial probit models. \citet{burgette2012trace} show that fixing the trace of the covariance matrix should be preferred over fixing a diagonal element, as in \citet{mcculloch2000bayesian} and \citet{imai2005bayesian}. Our model reparametrization satisfies the trace restriction and also naturally imposes parsimony, which is not embedded in the marginal data augmentation approach used by \citet{burgette2012trace}. 

Third, this is the first study that applies the multinomial probit model to real choice data with a large number of choice alternatives. Empirical applications of multinomial probit models have been limited to only a few choice alternatives. For instance, \citet{imai2005bayesian} consider six clothing detergent brands, \citet{mcculloch1994exact} and \citet{burgette2012trace} six margarine brands, \cite{piatek2017multinomial} two education levels and three occupation categories, and \citet{cripps2009parsimonious} five tests for cervical cancer. This limitation is of particular concern today, with the widespread availability of data on large choice sets. 

The outline of the remainder of this paper is as follows. Section~\ref{sec: mnp} discusses the model specification and Section~\ref{sec: scalable} introduces a scalable covariance matrix specification. Section~\ref{sec: inference} discusses prior specifications and the MCMC sampler. Section~\ref{sec: simulexe} conducts a numerical experiment to evaluate estimation accuracy, and Section~\ref{sec: application} applies the proposed methods to real consumer choice data sets. Section~\ref{sec: conclusion} concludes. 
\section{Multinomial probit model}\label{sec: mnp}

\subsection{Model specification}
Let $Y_i$ be an observable unordered random categorical variable
with support on the set $A_J = \{0,1,2,\dots,J\}$, with $J+1$ the number of choice alternatives, and $i=1,\dots,N$, with $N$ the number of individuals. Let $\tilde{Z}_i = (\tilde{z}_{i0},\dots,\tilde{z}_{iJ})^\top$ be a $(J+1)\times1$ vector of continuous random variables that can be interpreted as latent utilities, with 
\begin{align}
     Y_i(\tilde{Z}_{i}) &= \argmax_{j\in A_J} \tilde{z}_{ij}.
\end{align}
The latent utilities are modeled as
\begin{align}\label{eq: Ztil}
  \tilde{Z}_{i} &= \tilde{X}_i \tilde{\beta} +\tilde{\varepsilon}_{i}, \quad \tilde{\varepsilon}_i \sim N(0,\tilde{\Sigma}), 
\end{align}
where $\tilde{X}_i$ is a $(J+1)\times \tilde{K}$ matrix of observed regressors, $\tilde{\beta}$ is a vector of coefficients, and $\tilde{\varepsilon}_i$ is an independent normally distributed disturbance vector with covariance matrix $\tilde{\Sigma}$. The regressor matrix typically includes an intercept, a $k_d$-dimensional vector $x_{i,d}$ of individual-specific characteristics, and a $(J+1)\times k_a$ matrix $x_{i,a}$ of $k_a$ alternative-specific covariates, such that
\begin{align}\label{eq: Xtil}
    \tilde{X}_i = [I_{J+1} \quad x_{i,d}^\top \otimes I_{J+1} \quad x_{i,a}],
\end{align}
where $I_{J+1}$, denotes the identity matrix of dimensions $(J+1)\times(J+1)$.

\subsection{Identification}
The parameters $\tilde{\beta}$ and $\tilde{\Sigma}$ in the multinomial probit model specified in \eqref{eq: Ztil} are not identified \citep{bunch1991estimability}. There are two parameter identification problems. First, the location of the latent utilities is unidentified, since  $Y_i(\tilde{Z}_i+c) = Y_i(\tilde{Z}_i)$ for $c\in\mathbb{R}$. Second, the scale of the latent utilities is also unidentified, as $Y_i(c\tilde{Z}_i) = Y_i(\tilde{Z}_i)$ for $c\in\mathbb{R}^+$.

\subsubsection{Location}
A standard solution to the identification problem in the location is to difference the utilities with respect to a baseline category. Define choice category $j=0$ as the base category and define the differences in utilities as $Z_{ij}=\tilde{z}_{ij}-\tilde{z}_{i0}$. The dependent variable $Y_i$ equals
\begin{align}\label{eq: y}
    Y_i =  \left\{
    \begin{array}{ll}
        0 & \mbox{ if }  \max(Z_i)<0,\\
        j & \mbox{ if } z_{ij} = \max(Z_i)>0, 
    \end{array}
\right.
\end{align}
where $\max(Z_i)$ is the largest element of $Z_i$.

The utility model in \eqref{eq: Ztil} is transformed to differences in utilities by
\begin{align}\label{eq: Z}
    Z_i = T\tilde{Z}_i = {X}_i {\beta} +{\varepsilon}_{i}, \quad {\varepsilon}_i \sim N(0,{\Sigma}), 
\end{align}
with transformation matrix $T=\left[-\iota_J \quad I_J\right]$, the $J \times K$ transformed regression matrix $X_i=[I_{J} \quad x_{i,d}^\top \otimes I_{J} \quad Tx_{i,a}]$, and the $J \times J$ transformed covariance matrix $\Sigma=T\tilde{\Sigma}T^\top$. For the remainder of this paper we employ this location identification approach, and refer to $Z_i$ as utilities. Moreover, we define $Y=(Y_1,\dots,Y_N)^\top$, $Z=(Z_1^\top,\dots,Z_N^\top)^\top$, and $X=(X_1^\top,\dots,X_N^\top)^\top$.
 
\subsubsection{Scale}
There are multiple solutions to the unidentified scale in the latent utilities, but they all impose a constraint on the covariance matrix $\Sigma$. \citet{mcculloch2000bayesian} develop a prior that fixes the (1,1) element of $\Sigma$ to be equal to one. \citet{burgette2012trace} argue that the assignment of a choice alternative to the unit variance can have a large effect on the posterior choice probabilities. They propose to fix the trace of the covariance matrix instead of fixing one of its elements. We follow this approach and  restrict the trace of the covariance matrix $\Sigma$ in \eqref{eq: Z} to be equal to $J$.
\section{Scalable reparametrization}\label{sec: scalable}
The total number of unique parameters in the covariance matrix, $J(J+1)/2$, grows quadratically with $J$. To reduce the dimension of the parameter space, Section~\ref{sec:factor} specifies a factor structure for $\Sigma$. Section~\ref{sec:trace} introduces a reparametrization of the factor structure that imposes a trace restriction on the covariance matrix.

\subsection{Factor structure}\label{sec:factor}
Denote as ${\gamma}$ a $J\times q$ matrix with $q<J$, and as $D$ a diagonal matrix with positive diagonal elements $d=(d_1,\dots,d_J)$. We model $\Sigma$ as
\begin{align}\label{eq: sigmafactor}
    {\Sigma} = {\gamma} {\gamma}^\top + {D^2}.
\end{align}
For the purpose of identification, the upper triangular elements of $\gamma$  are fixed at zero \citep{geweke1996measuring}. In this factor covariance structure, the total number of parameters that characterise the covariance matrix is $n = J(q+1)-q(q-1)/2$, which implies that for a given value of $q$, the number of parameters grows linearly with $J$. 

\subsection{Trace restriction}\label{sec:trace}
A major challenge in the factor covariance structure is the implementation of the identifying restriction $\text{trace}(\Sigma) = J$, which implies 
\begin{align}\label{eq: trace}
   \text{trace}(\Sigma)=  \sum_{j=1}^J[(\sum_{k=1}^{\text{min}(q,j)} \gamma_{jk}^2)+d_j^2] = J,
\end{align}
where the scalar $\gamma_{jk}$ denotes the element in row $j$ and column $k$ in $\gamma$, while the scalar $d_j$ is the $j^{\text{th}}$ element in $d$.  

The trace restriction implies that by construction, the elements of $\gamma$ and $d$ need to be constrained to the surface of an $n-$sphere of radius $\sqrt{J}$. To show this, define the $n$-dimensional vector $\psi$ as
\begin{align}\label{eq: psi}
    \psi = (\psi_1,\dots,\psi_n)^\top= \left(d^\top,\text{vech}(\gamma)^\top\right)^\top,
\end{align}
where vech$(\gamma) =\left(\gamma_{1:J,1}^\top,\dots,\gamma_{q:J,q}^\top\right)^\top$ and $\gamma_{k:J,k}=\left(\gamma_{kk},\dots,\gamma_{Jk}\right)^\top$. The trace restriction in \eqref{eq: trace} is equivalent to the spherical restriction $\sum_{l=1}^n\psi_l^2 =J$ on $\psi$, from which follows that the elements of $\psi$ are restricted to the $n-$dimensional spherical space $\mathbb{S}^n = \{\psi: \sum_{l=1}^n\psi_l^2 =J\}$.

As one must guarantee the {spherical} restriction, which involves all the elements in $\psi$, estimation of $\psi$ is challenging. To avoid direct implementation of this restriction, we exploit the fact that $\psi$ is restricted to $\mathbb{S}^n$, which naturally allows for a spherical transformation from $\psi$ into an  $(n-1)$-dimensional vector of angles $\kappa = \left(\kappa_1,\dots,\kappa_{n-1}\right)^\top$ and the radius $\sqrt{J}$, with $\kappa\in\mathbb{A}$ and $\mathbb{A} = [0,\pi)^{n-2}\times[0,2\pi)$.

The spherical transformation reparametrises $\psi_l$ as
\begin{equation}\label{EQ:inverse_function}
  \psi_l({\kappa},J) = \begin{cases}
      \sqrt{J}\cos\kappa_1 & \text{for $l=1$},\\
    \sqrt{J}\cos\kappa_l\prod_{j = 1}^{l-1}\sin\kappa_j &   \text{for $1<l<n$},\\
    \sqrt{J}\prod_{j = 1}^{l-1}\sin\kappa_j & \text{for $l=n$}.
  \end{cases}
\end{equation}
The main advantage of this transformation is that inference on the parameter space $\mathbb{A}$ is a more accessible problem, as no {spherical} restriction is required for $\kappa$. Moreover, the inverse function of the transformation is available in closed-form,
\begin{equation}\label{EQ:transformation}
  \kappa_l({\psi}) = \begin{cases}
      \arccos\left[\psi_l\left(\sum_{j=l}^{n}\psi_j^2\right)^{-\frac{1}{2}}\right] & \text{for $l<n-1$},\\
      \arccos\left[\psi_l\left(\sum_{j=l}^{n}\psi_j^2\right)^{-\frac{1}{2}}\right] &   \text{for $\{l =n-1 \land \psi_n\ge0\}$},\\
    2\pi-\arccos\left[\psi_l\left(\sum_{j=l}^{n}\psi_j^2\right)^{-\frac{1}{2}}\right] & \text{for $\{l =n-1 \land \psi_n<0\}$}.
  \end{cases}
\end{equation}    
Notice from \eqref{EQ:transformation} that $\kappa_l(c{\psi}) = \kappa_l({\psi})$, for any positive scalar $c$. Setting $c=\sum_{l=1}^n\psi_l^2=\text{trace}(\Sigma)$, shows that $\kappa_l$ is a function of the scale of $\psi_l$ relative to the other elements in $\psi$, rather than the trace of $\Sigma$. 
\section{Bayesian estimation}\label{sec: inference}
This section develops a Bayesian method for estimating the identified parameters $(\beta,\kappa)$, subject to the trace restriction in \eqref{eq: trace}. 
The {density} of interest is the augmented posterior \begin{equation}\label{eq:posterior}
p\left(\beta,\kappa,Z|Y,X,B,\theta\right)\propto p(Y|Z)p(Z|X,\beta,\kappa)p(\beta|B)p(\kappa|\theta),
\end{equation}
where we assume the priors of $\beta$ and $\kappa$ to be independent, with corresponding hyperparametes $B$ and $\theta$. The prior on the coefficients $\beta$ is standard and specified as
\begin{align}\label{eq:priorbeta}
    \beta|B \sim N(0,B^{-1}).
\end{align}
We develop the prior choice for $p(\kappa|\theta)$ in the next section.

\subsection{Prior for angular coordinates}\label{sec: angles}
Developing prior beliefs directly on the angular coordinates $\kappa$ is challenging, because these coordinates lack interpretation in relation to the covariance matrix $\Sigma$. On the other hand, because $\psi$ directly determines the covariance matrix $\Sigma$ via \eqref{eq: sigmafactor} and \eqref{eq: psi}, the implications that a prior on $\psi$ has on the prior beliefs on $\Sigma$ are well understood. At the same time, choosing a prior on $\psi$ imposes the prior on $\kappa$ required in \eqref{eq:posterior}. 

Therefore, we first select a prior $p\left(\psi|\theta\right)$. This prior implies $p(\kappa|\theta)$, which is required for parameter estimation, and $p(\Sigma|\theta)$. Because the elements of $\kappa$ and $\psi$ do not have a monotonically increasing relationship, analytical derivation of $p(\kappa|\theta)$ is challenging. Instead, we construct a parametric approximation to $p\left(\kappa|\theta\right)$ from a flexible parametric density class $\mathcal{P}$ with elements $\tilde{p}\left({\kappa}|{\lambda}\right)$ indexed by $\lambda\in\Lambda$. 

The approximating prior density $\tilde{p}({\kappa}|\hat{\lambda})$ is calibrated by minimizing an estimate of the Kullback-Leibler divergence $\text{KL}\left[p\left(\kappa|\theta\right)||\tilde{p}\left({\kappa}|\lambda\right)\right]$, with respect to $\lambda$. Specifically, we minimize 
\begin{align}\label{Eq:KL}
\widehat{\text{KL}}\left[p\left(\kappa|\theta\right)||\tilde{p}\left({\kappa}|\lambda\right)\right]=\frac{1}{M}\sum_{m=1}^M\log\left(p(\kappa^{[m]}|\theta)\right)-\frac{1}{M}\sum_{m=1}^M\log\left(p(\kappa^{[m]}|\lambda)\right),
\end{align}
where the first term can be ignored in the minimization problem, and the second term is computed using $M$ draws $\{\kappa^{[m]}\}_{m=1}^{M}$.
These draws are produced by generating from the prior distribution $p({\psi}|\theta)$, and then transforming into draws from the prior $p\left({\kappa}|\theta\right)$. Algorithm~\ref{alg: simulation} outlines the steps of the optimization process.

\begin{algorithm}
\caption{Prior calibration on the angular coordinates}\label{alg: simulation}
\begin{algorithmic}[1]
\State Set hyperparameters $\theta$ in $p({\psi}|\theta)$
\State Generate $M$ draws $\psi^{[m]}\sim p({\psi}|\theta)$
\State Transform $\psi^{[m]}$ to $\kappa^{[m]}$ using \eqref{EQ:transformation}
\State Calculate $\hat{\lambda}$ as the minimizer of \eqref{Eq:KL} using $\{\kappa^{[m]}\}_{m=1}^{M}$ 
\State Construct $\tilde{p}(\kappa|\hat{\lambda})$
\end{algorithmic}
\end{algorithm}

Once calibrated, we can use the prior $\tilde{p}(\kappa|\hat{\lambda})$ for inference, by plugging it into \eqref{eq:posterior} instead of $p\left({\kappa}|\theta\right)$.
Key to accurate approximation of $p\left({\kappa}|\theta\right)$ is the selection of a flexible parametric density class  $\mathcal{P}$. Here we use $\tilde{p}(\kappa|{\lambda})=\prod_{l=1}^{n-1}\tilde{p}(\kappa_l|{\lambda}_l)$ with
\begin{equation}\label{EQ:xi_density}
\tilde{p}(\kappa_l|\lambda_l) = \phi_1\left\{t_{\eta_l}\left[\frac{G\left(\kappa_l\right)-\mu_l}{\tau_l}\right]\right\}t_{\eta_l}'\left[\frac{G\left(\kappa_l\right)-\mu_l}{\tau_l}\right]\frac{1}{\tau_l}G'\left(\kappa_l\right),
\end{equation}
where $G\left(\kappa_l\right) = \Phi_1^{-1}\left(\frac{\kappa_l}{\pi}\right)$ for $l<n-1$, $G\left(\kappa_l\right) = \Phi_1^{-1}\left(\frac{\kappa_l}{2\pi}\right)$ for $l=n-1$, $G'()$ is the derivative of $G()$, $t_{\eta}()$ is the \cite{yeo2000new} transformation, $t_{\eta}'()$ its first derivative, while $\phi_1()$ and $\Phi^{-1}_1()$ denote the density and inverse distribution function of a standard normal variable, respectively.

The density function in \eqref{EQ:xi_density} is capable of accurately approximating the margins of the prior $p(\kappa|\theta)$, which in turn results in accurate approximation to the implied prior $p(\Sigma|\theta)$, as we will show in Section~\ref{sec: demonstration}. Moreover, the evaluation of the density is computationally efficient, which increases the speed of the sampling algorithm discussed in Section~\ref{sec:scheme}. Therefore, we select this density class over generally more computationally involved non-parametric alternatives. For details on the properties and construction of this distribution we refer to Appendix~\ref{App:FlexibleDistribution}.

\subsection{Choice of prior}\label{sec: demonstration}
To set the prior $p(\kappa|\theta)$, the practitioner needs to specify a prior on the vector ${\psi}$. Although simulation from $p(\psi|\theta)$ is required for Algorithm~\ref{alg: simulation}, this prior density does not have to be available in closed-form. 

We propose the following characterisation of the prior distribution $p(\psi|\theta)$,
\begin{align}
    \psi &= \frac{\sqrt{J}}{\lVert\ddot{\psi}\lVert}\ddot{\psi},\label{eq: psiprior0}\\
    p(\ddot{\psi}|\theta) &= \prod_{j=1}^{J}\left[p({\ddot{d}}_j|\nu,s)\prod_{k=1}^{\text{min}(q,j)}p({\ddot{\gamma}}_{jk}|\sigma^2_{\gamma})\right],\label{eq: psiprior1}\\
    {\ddot{\gamma}_{jk}|\sigma_{\gamma}^2}&\sim N({\mu_{\gamma}},\sigma_{\gamma}^2),\label{eq: psiprior3}\\
    {\ddot{d}}_j^2|\nu,s&\sim\text{Inverse-Gamma}\left(\nu,s\right),\label{eq: psiprior4}
\end{align}
where $\nu$ and $s$ denote the shape and rate parameters of the Inverse-Gamma distribution. 

This choice of prior links the trace-restricted parameters in $\psi$ to an unrestricted parameter vector $\ddot{\psi}=(\ddot{d}^\top,\text{vech}(\ddot{\gamma})^\top)^\top$. As a result, we induce a prior on the restricted parameters $\psi$ by selecting a prior for the unrestricted parameters $\ddot{\psi}$. For $\ddot{\psi}$ any prior that fits the type of factor structure in \eqref{eq: sigmafactor} may be employed. We use a particularly well-established prior in the factor literature, that assumes a normal distribution on $\ddot{\gamma}_{jk}$ and an Inverse-Gamma distribution for $\ddot{d}^2_j$ (see for instance \cite{lopes2014modern} and references therein).  

Since $\psi$ is a function of the scale of the elements of $\ddot{\psi}$ relative to the norm of $\ddot{\psi}$, the location of $p(\psi|\theta)$ is unidentified. To solve this, we anchor the mean of the Inverse-Gamma prior in \eqref{eq: psiprior4} at one by setting $s = \nu-1$. Thus, the hyperparameters for the prior are $\theta = \left(\mu_{\gamma},\sigma_\gamma,\nu\right)^\top$.

\subsubsection{Accuracy of the approximating prior}
This section assesses the accuracy of the approximating density for the choice of prior in the previous section. We focus on the parameter space of most interest to the practitioner, the covariance matrix $\Sigma$, which is implied by $\tilde{p}(\kappa|\hat{\lambda})$. Specifically, we focus on assessing the accuracy of $\tilde{p}(\Sigma_{2,2}|\hat{\lambda})$ at replicating ${p}(\Sigma_{2,2}|\theta)$, and the accuracy of $\tilde{p}(\rho_{2,3}|\hat{\lambda})$ at replicating ${p}(\rho_{2,3}|\theta)$ with $\rho_{2,3} =\frac{\Sigma_{2,3}}{\sqrt{\Sigma_{2,2}\Sigma_{3,3}}}$, for $J=6$, $q=1$, and $\theta=(0,1,5)^\top$. All the results in this section also hold for the other elements in $\Sigma$ and for any number of choice alternatives $J$.

The prior ${p}(\Sigma_{2,2}|\theta)$ is constructed via simulation. First, we generate draws from the prior $p(\kappa|\theta)$, then transform them into draws for $\Sigma$, and finally construct a kernel density estimator of $p(\Sigma_{2,2}|\theta)$. Similarly, we use Algorithm~\ref{alg: simulation} to calibrate the approximating prior $\tilde{p}(\kappa|\hat{\lambda})$, from which we also obtain a kernel density estimator of $\tilde{p}(\Sigma_{2,2}|\hat{\lambda})$ via simulation. We construct $p(\rho_{2,3}|\theta)$ and $\tilde{p}(\rho_{2,3}|\hat{\lambda})$ in the same way. 

\begin{figure}[tb!]
\caption{Approximating prior density for $\Sigma_{2,2}$}
\centering
\includegraphics[width=\textwidth]{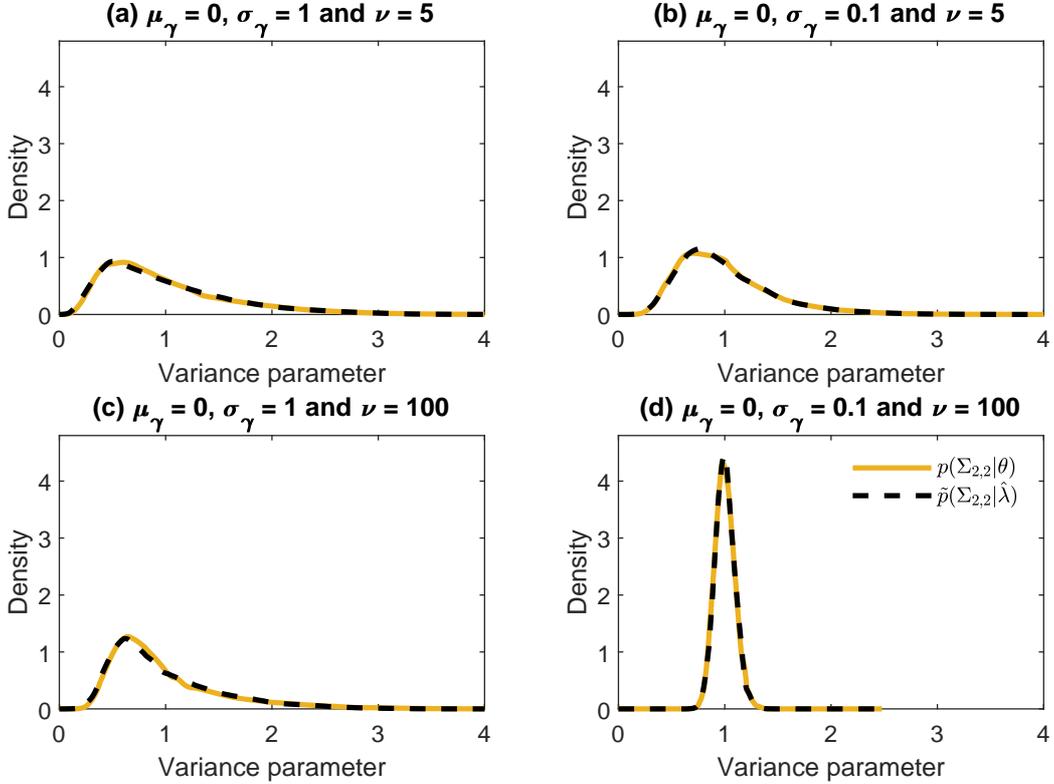}
 \fnote{This figure shows the implied prior variance densities for the multinomial probit model with a factor structure. The yellow solid line corresponds to the implied prior $p(\Sigma_{2,2}|\theta)$ and the black dashed line corresponds to its approximation $\tilde{p}(\Sigma_{2,2}|\hat{\lambda})$, and the panels correspond to different values for $\theta$ with $q=1$.}
\label{Fig:tager_vs_calibrated_sigma}
\end{figure}

\begin{figure}[tb!]
\caption{Approximating prior density for $\rho_{2,3}$}
\centering
\includegraphics[width=\textwidth]{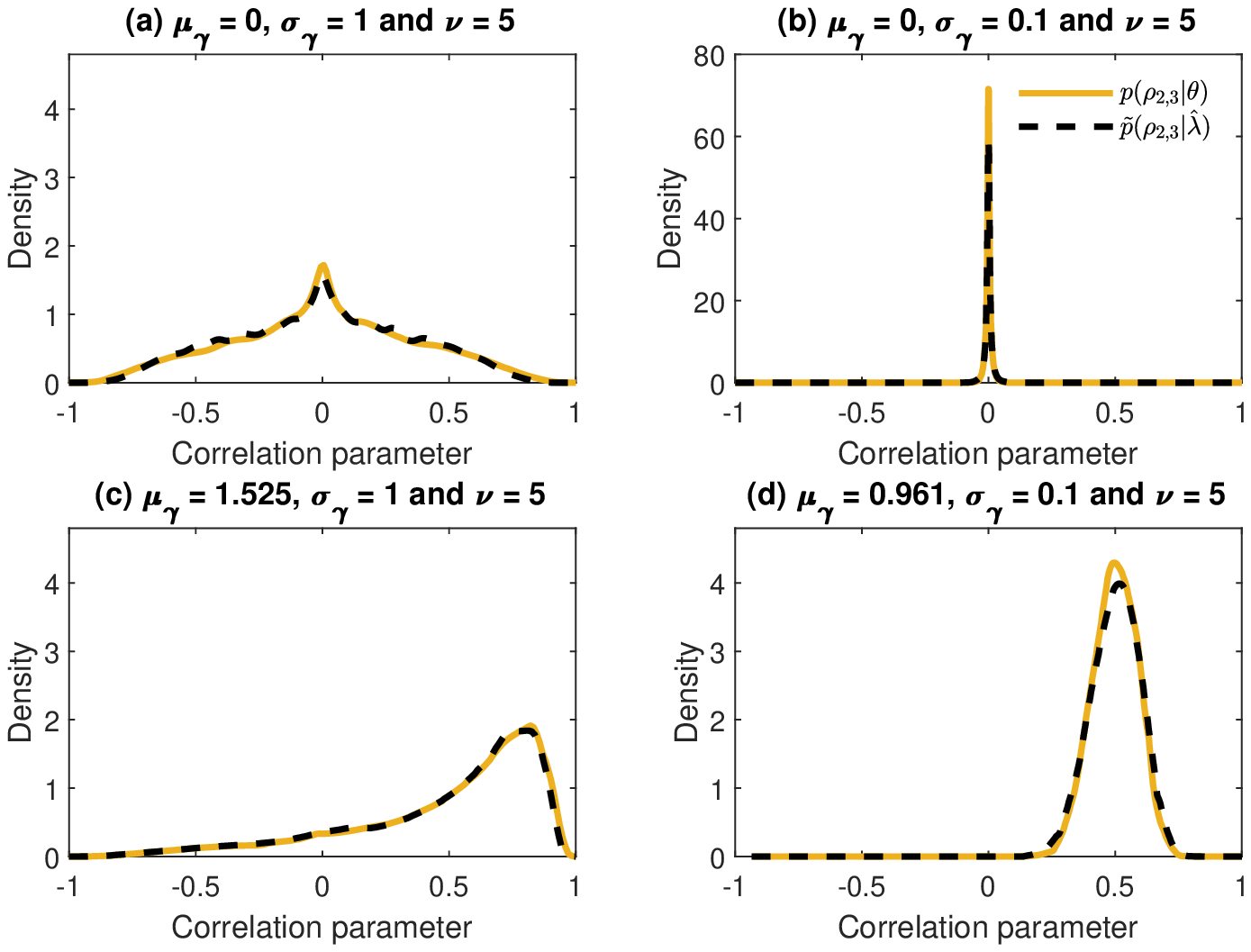}
 \fnote{This figure shows the implied prior correlation densities for the multinomial probit model with a factor structure. The yellow solid line corresponds to the implied prior $p(\rho_{2,3}|\theta)$ and the black dashed line corresponds to its approximation $\tilde{p}(\rho_{2,3}|\hat{\lambda})$, and the panels correspond to different values for $\theta$ with $q=1$.}
\label{Fig:tager_vs_calibrated_corr}
\end{figure}

In Panel (a) of Figure~\ref{Fig:tager_vs_calibrated_sigma}, the yellow solid line and the black dashed line represent the implied prior densities $p(\Sigma_{2,2}|\theta)$ and $\tilde{p}(\Sigma_{2,2}|\hat{\lambda})$, respectively. The approximating prior is an accurate representation of the prior. The remaining panels in Figure~\ref{Fig:tager_vs_calibrated_sigma} demonstrate that the approximating prior remains accurate for alternative hyperparameter values. A similar result is obtained when we compare the implied priors $p(\rho_{2,3}|\theta)$ and $\tilde{p}(\rho_{2,3}|\hat{\lambda})$, as shown in  Figure~\ref{Fig:tager_vs_calibrated_corr}.

\begin{figure}[tb!]
\caption{Approximating prior densities with four factors}
\centering
\includegraphics[width=\textwidth]{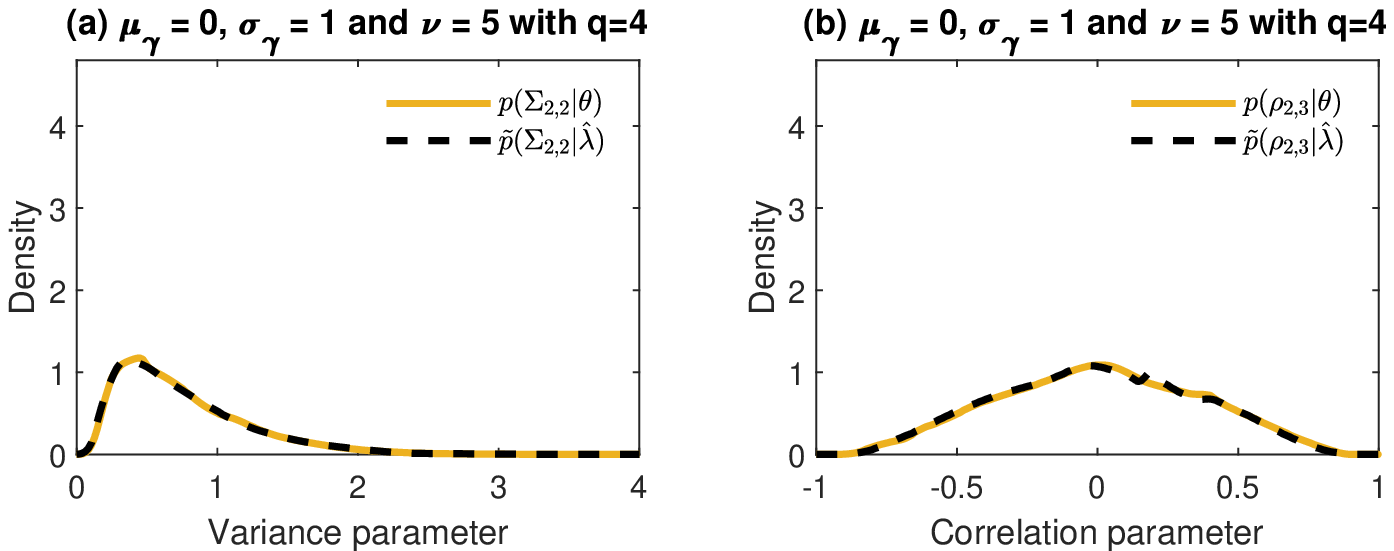}
 \fnote{This figure shows the implied prior variance (panel a) and correlation (panel b) densities for the multinomial probit model with a factor structure. The yellow solid line corresponds to the implied prior and the black dashed line corresponds to its approximation.}
\label{Fig:tager_vs_calibrated_q}
\end{figure}

Additionally, Figure~\ref{Fig:tager_vs_calibrated_q} shows that also for a larger number of factors $q=4$ the approximating prior is accurate. 

\subsubsection{Hyperparameters}
The user of the prior proposed in \eqref{eq: psiprior0}-\eqref{eq: psiprior4} only has to set the hyperparameters $\theta$. Here we discuss the impact that these hyperparameters have on the implied prior for $\Sigma$. The parameters $\sigma_{\gamma}$ and $\nu$ jointly control the dispersion of the variances around their prior mean of one, and $\sigma_{\gamma}$ also governs the variance of the correlations around their prior mean. 

To illustrate this, Panels (b) and (d) in Figure~\ref{Fig:tager_vs_calibrated_sigma} show that, for the small value of $\sigma_{\gamma}=0.1$, a larger value of $\nu$ makes the prior on the diagonal elements of $\Sigma$ tighter around one. In Panels (a) and (c), we observe that for large values of $\sigma_{\gamma}$ the hyperparameter $\nu$ has little effect on the prior. On the other hand, comparing Panel (a) to (b), and Panel (c) to (d), indicates that for smaller values of $\sigma_{\gamma}$ the impact of $\nu$ on the prior is more pronounced.

Panels (a) and (b) in Figure~\ref{Fig:tager_vs_calibrated_corr} show that $\sigma_\gamma$ governs the prior variance on the underlying correlations of $\Sigma$, with larger values for $\sigma_\gamma$ associated with a prior with larger variance. The parameter $\nu$ does not affect the implied prior on the correlations. So a small value for $\sigma_{\gamma}$ shrinks the correlations towards their prior means. A comparison of Panels (a) and (b) to Panels (c) and (d) indicates that the prior mean of $\Sigma$ equals $I_J$ for $\mu_{\gamma}=0$, because the correlations have a prior mean of zero, and it equals an equicorrelated matrix when $\mu_{\gamma}\ne0$.

The implied prior for $\Sigma$ is also sensitive to the number of factors considered. 
Figure~\ref{Fig:tager_vs_calibrated_q} shows how these implied prior densities change when the number of factors is set to $q=4$. Increasing the number of factors has a small effect in the priors for both the variance and correlation parameters. 

For the remainder of this paper we consider two choices for $\mu_{\gamma}$. The first choice, $\mu_{\gamma}=0$, sets the prior mean of the covariance matrix $\Sigma$ equal to the identity matrix, which is a standard MNP prior choice as we will discuss in the next section. However, $\Sigma$ is the covariance matrix of the differences in utilities $Z$, for which an identity covariance matrix does not necessarily imply a symmetric correlation structure for the untransformed utilities $\tilde{Z}$. Therefore, the second choice, $\mu_{\gamma}=\mu_\gamma^*$, sets the prior mean of $\Sigma$ equal to the equicorrelated covariance matrix $\frac{1}{2}(I_J + \iota_J \iota_J^\top)$. This follows \citet{geweke1994alternative}, who shrink the covariance matrix of the untransformed utilities $\tilde{\Sigma}$ to an identity matrix, which is equivalent to shrinking $\Sigma$ to this equicorrelated matrix. The value $\mu_\gamma^*$ that produces the equicorrelated matrix above can be computed as in Appendix \ref{sec:equicorr}, for any given values of $\sigma_{\gamma}$, $\nu$ and $q$.

For the remaining hyperparameters, we use the values $\sigma_{\gamma}=1$, $\nu=5$, and $q=1$. These hyperparameter values produce prior densities for the variances that have close to zero probability mass at zero. They also imply prior densities for the correlations that have low probability mass at one and minus one. These properties make the method computationally stable, and guarantee that extreme values for the variances and correlations are the result of a strong signal in the data instead of highly uninformative priors.

\subsubsection{Comparison to existing MNP prior specifications}
This section compares our prior assumptions on $\Sigma$ to two well-known alternatives in the MNP literature. The first, from here on in referred to as MNP-MPR, is proposed by \citet{mcculloch2000bayesian}, who set
\begin{align}\label{eq:MRprior}
  \Sigma &= \begin{bmatrix}
    1 & \gamma^\top \\ \gamma & \Phi+\gamma \gamma^\top
    \end{bmatrix},
    \quad \gamma \sim N(0,\tau I_{J-1}),\quad \Phi^{-1}\sim \text{Wishart}(\delta,C),
\end{align}
with degrees of freedom $\delta=J+3$ and scale matrix $C=(\delta-J)(1-\tau)I_{J-1}$, where $\tau=\frac{1}{8}$. The second alternative, from here on in MNP-BN, was proposed in \citet{burgette2012trace}, who specify
\begin{align}\label{eq:BNprior}
    \Sigma = \ddot{\Sigma}/\text{trace}(\ddot{\Sigma}),\text{ with } \ddot{\Sigma}\sim \text{Inverse-Wishart}(s,S),
\end{align}
with degrees of freedom $s=J+3$ and scale matrix $S=I_J$.

\begin{figure}[tb!]
\caption{Prior densities for variance and correlation parameters}
\centering
\includegraphics*[width=\textwidth,trim = 0 0 0 0]{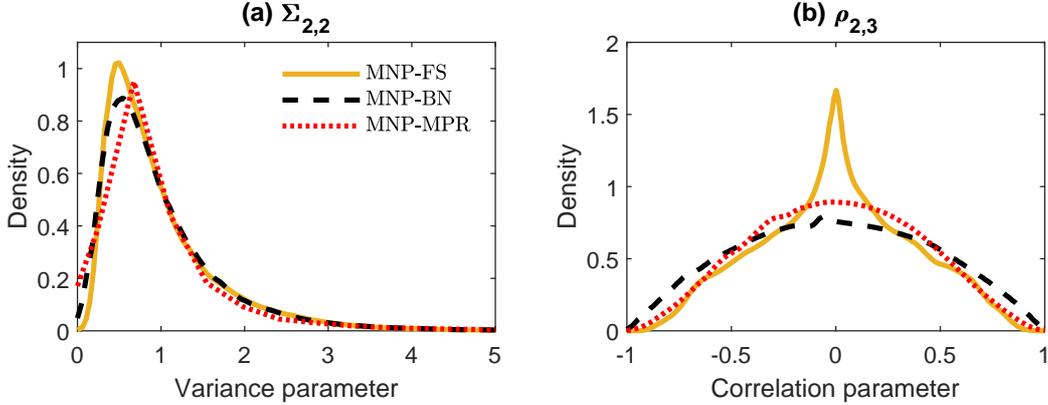}
 \fnote{Panels (a) and (b) show the implied prior densities of $\Sigma_{2,2}$ and $\rho_{2,3}$, respectively. The yellow solid, red dashed and black dotted lines corresponds to the MNP-FS, MNP-BN and MNP-MNR approaches, respectively.}
\label{fig:priortune}
\end{figure}

Panel (a) in Figure~\ref{fig:priortune} compares the implied prior on $\Sigma_{2,2}$ for our proposed multinomial probit model with factor structure, from here on in referred to as MNP-FS, to MNP-BN and MNP-MPR. For MNP-FS we consider $q=1$ and $\theta=(0,1,5)^\top$. The three densities are positively asymmetric and most of their probability mass lies between zero and four. Aside from the fact that the MNP-FS density has low mass at zero, all three prior densities have a similar shape.

Panel (b) in Figure~\ref{fig:priortune} compares the implied priors on the correlation element $\rho_{2,3}$. The MNP-BN and MNP-MPR priors assign more probability mass to the edges of the support, while the MNP-FS prior is {slightly} tighter and has a spike at zero. Although not visible in the figure, the MNP-FS prior still assigns enough probability mass at extreme values of the support, to allow for accurate estimation of correlations whose true parameter values are extreme. 

\subsection{Sampling scheme}\label{sec:scheme}
To construct the posterior in \eqref{eq:posterior}, we employ the following MCMC sampling scheme:\\
\ \\
$\underline{\text{Sampling Scheme}}$\\
\ \ \hspace{2cm} Step 1: Generate from $\beta|Z,\kappa,X,B$.  \\ 
\ \ \hspace{2cm} Step 2: Generate from $Z|\beta,\kappa,Y,X$.\\
\ \ \hspace{2cm} Step 3: Generate from $\kappa|Z,\beta,X,\theta$.\\
\ \\
Steps 1 and 2 are standard Gibbs sampling steps, see for instance \citet{mcculloch1994exact}. These steps require one to transform from $\kappa$ to $\Sigma$, which can be easily achieved following the instructions outlined in Table~\ref{tab:trans}. 

\begin{table}[tb!]
  \centering \small
  \caption{Transformations between angles and covariance parameters}
  \begin{threeparttable}
			\resizebox*{\textwidth}{!}{
			\begin{tabular}{ccccccc}
				\toprule\toprule
				\multicolumn{7}{c}{{ \textbf{(a) Transforming from} $\kappa$ \textbf{to} $\psi$ \textbf{and} $\Sigma$}}\\
				\midrule
				  Angles && $\Rightarrow$ && Cov parameters &$\Rightarrow$& Cov matrix  \\ \cline{1-1}\cline{5-5}\cline{7-7}

   $\kappa$      &&    $
				 \psi_l = \begin{cases}
				 	\sqrt{J}\cos\kappa_1                                & \text{for\ } l=1,   \\
				 	\sqrt{J}\cos\kappa_l\prod_{j = 1}^{l-1}\sin\kappa_j & \text{for\ } 1<l<n, \\
				 	\sqrt{J}\prod_{j = 1}^{l-1}\sin\kappa_j             & \text{for\ } l=n
				 \end{cases}
				 $                           &&  $\psi = \left(d^\top,\text{vech}(\gamma)^\top\right)^\top$&$\Sigma = \gamma\gamma^\top+D^2$&$\Sigma$\\
    \midrule
				\multicolumn{7}{c}{{ \textbf{(b) Transforming from} $\psi$ \textbf{to} $\kappa$}} \\ \midrule
			 Angles && $\Leftarrow$ &&Cov parameters&&   \\\cline{1-1}\cline{5-5}
   $\kappa$      &&\multicolumn{2}{l}{    $  \kappa_l = \begin{cases}
				\arccos\left[\psi_l\left(\sum_{j=l}^{n}\psi_j^2\right)^{-\frac{1}{2}}\right] & \text{for $l<n-1$},\\
				\arccos\left[\psi_l\left(\sum_{j=l}^{n}\psi_j^2\right)^{-\frac{1}{2}}\right] &   \text{for $\{l =n-1 \land \psi_n\ge0\}$},\\
				2\pi-\arccos\left[\psi_l\left(\sum_{j=l}^{n}\psi_j^2\right)^{-\frac{1}{2}}\right] & \text{for $\{l =n-1 \land \psi_n<0\}$}
				\end{cases}
				 $                           }&$\psi = \left(d^\top,\text{vech}(\gamma)^\top\right)^\top$&&\\ 
				 \bottomrule \bottomrule
			\end{tabular}
		}
\begin{tablenotes}
\footnotesize
\item Depiction of transformations to and from the parameters $\kappa$ to covariance matrix $\Sigma$.\\ Here, $D$ denotes the diagonal matrix with diagonal elements $d$.
\end{tablenotes}
\end{threeparttable}
  \label{tab:trans}
\end{table}

For step 3 we employ a random walk Metropolis-Hastings sampler. At the beginning of each iteration, the elements of $\kappa$ are randomly assigned to groups of five elements. The groups are then sampled, one group conditional on the other, with the $5$-dimensional proposal density equal to the product of five independent truncated univariate normals. The variances of the  proposal densities are set adaptively to target acceptance rates between $15\%$ and $30\%$. 

The random assignment into groups, plus the parameter-specific adaptive steps, allow one to target parameter-specific acceptance rates without having to sample each element of $\kappa$ one at a time. \citet{roberts2009examples} provide more details on adaptive MCMC, and \citet{smith2015copula} provides an illustration on random allocation within MCMC. Appendix~\ref{sec:mcmc} discusses the sampling steps in more detail. 

{Although the reparametrization of $\Sigma$ into $\kappa$ results in a non-conjugate sampling step, the added computational cost is small. This is because Step 2 - also required for the competing MNP estimation approaches - is the most time consuming of the sampling scheme, especially for large $J$. Section~\ref{sec: application} illustrates this in the simulation experiment.}

{Potential alternatives to our reparametrization that produce a conjugate Gibbs sampling scheme may complicate the inference of the factor structure. For instance, the factor structure in \eqref{eq: sigmafactor} could potentially be sampled by augmenting the parameter space with a set of latent factors as in \citet{lopes2014modern}, in combination with the marginal data augmentation approach in \cite{burgette2012trace} to impose the trace restriction. However, as discussed in \cite{piatek2017multinomial}, measurement of latent factors in an MNP model without extra data or a priori knowledge of the factor loadings $\gamma$ is challenging.}

\section{Numerical experiment}\label{sec: simulexe}
This section presents a numerical experiment to assess the accuracy of the parameter estimates in the proposed multinomial probit model. First we describe the simulation design, second we discuss the results, and finally we examine how sensitive the fitted probabilities are to different base category specifications.

\subsection{Design}\label{sec:design}
We generate a data set from the data generating process specified in \eqref{eq: y} and \eqref{eq: Z}. The data closely matches the empirical application in Section~\ref{sec:applarge}, with the number of discrete choices $J+1 = 50$ and number of observations $N = 5000$. The elements of the vector $x_{i,a}$ are independently generated from normal distributions with corresponding mean $\mu=0$ and variance $\sigma^2=1$, and can be interpreted as the logarithm of the prices of the choice categories. We do not include individual-specific characteristics $x_{i,d}$. 

The true parameter vector $\beta_0$ consists of $J$ intercepts drawn independently from normal distributions with $\mu=0$ and $\sigma^2=\sqrt{0.5}$, and the coefficient for $x_{i,a}$ {which is} fixed at -0.7. The true covariance matrix $\Sigma_0$ is set by drawing $\tilde{\Sigma}_0$ from the $\text{Inverse-Wishart}(S,J+3)$ distribution, where the scale matrix $S$ has ones on the diagonal and 0.5 as the common off-diagonal element. We set $\Sigma_0=J\tilde{\Sigma}_0/\text{trace}(\tilde{\Sigma}_0)$.

We apply our method, MNP-FS, to the generated dataset. For the purpose of comparison we also implement the MNP-BN and MNP-MPR approaches. For our model, we set $\theta=(0,1,5)^\top$ with $q=1$ in the prior for $\kappa$. Setting the number of factors to one substantially reduces the number of covariance parameters to be estimated, especially with 50 choice alternatives. 
The implied prior densities at these parameter values are discussed in detail in Section~\ref{sec: demonstration}. 

We use the prior for the coefficients specified in \eqref{eq:priorbeta} with $B^{-1}=0.1I_K$ for all models. This is a rather uninformative prior, given that the covariates are scaled to a variance of one in the sampler. Since settings with large choice sets are vulnerable to numerical instabilities, we avoid the use of improper prior specifications. 

The posterior results are based on 200,000 iterations of the MCMC samplers, from which the first 100,000 are discarded. 

\subsection{Results}\label{sec:simulresults}
\begin{figure}[tb!]
\caption{Posterior mean parameters in numerical experiment}
\centering
\includegraphics*[width=\textwidth,trim = 0 0 0 0]{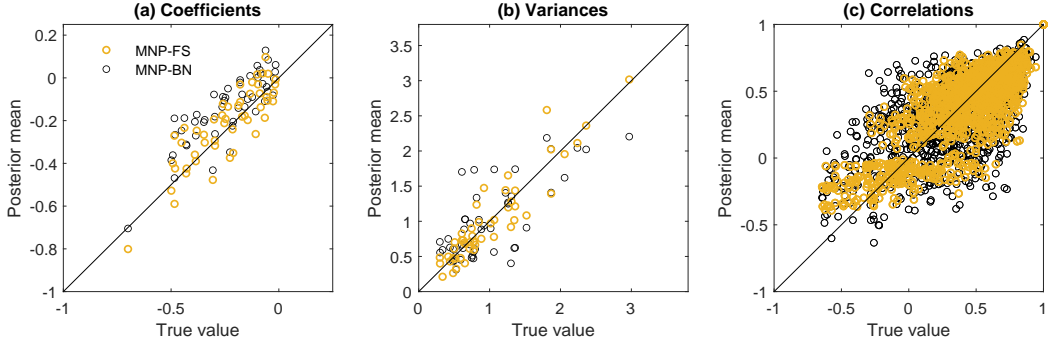}
 \fnote{{This figure presents the estimated posterior means from the MNP-FS} (yellow circles) and {MNP-BN}(black circles) approaches, for the coefficients $\beta$ in Panel (a), and the variances and correlations of the latent utilities in $\Sigma$ in Panel (b) and (c), respectively.}
\label{fig:simul}
\end{figure}

Figure~\ref{fig:simul} compares the true parameter values of $\beta_0$ and $\Sigma_0$ ($x-$axis) against the corresponding posterior mean estimates ($y-$axis). The yellow and black circles correspond to MNP-FS and MNP-BN, respectively. The closer the circles lie to the 45 degree diagonal line, the closer the posterior means are to the true parameter values. 

Panel (a) in Figure~\ref{fig:simul} {shows} that the MNP-FS approach provides more accurate estimates of $\beta_0$ than MNP-BN. This is also the case for the diagonal elements of $\Sigma_0$ in Panel (b). Finally, the results for the posterior mean estimates of the underlying correlations of $\Sigma_0$ are presented in panel (c) of Figure~\ref{fig:simul}. {The correlation estimates from MNP-FS are clearly the most tightly scattered around the diagonal line, and as such the most accurate.} {This result is particularly striking given the fact that the MNP-FS is the one imposing the most restrictive covariance structure.} Appendix~\ref{sec:addnum} shows that the comparison between MNP-FS and MNP-MPR, result in similar conclusions.

{The error measures reported in} Table~\ref{tab:errormeasures} confirm the {conclusions above}. The smallest root mean squared error (RMSE) of the posterior mean estimates for the variances and correlations correspond to the {MNP-FS} specification. The same conclusion is reached when comparison is conducted in terms of the mean absolute error (MAE), confirming that our approach allows for more accurate estimation of the elements in the covariance matrix. For the coefficients, the RMSE and MAE values from MNP-MPR are the smallest, closely followed  by MNP-FS.

{The computational costs of the methods are similar. Since MNP-FS requires several MH steps for the generation of $\kappa$, it takes $0.06$ seconds more per sample iteration than MNP-BN and MNP-MPR, which need on average $0.36$ and $0.35$ seconds for each sample iteration in this simulation exercise.}

    \begin{table}[tb!]
  \centering \small
 \caption{Error measures in the numerical experiment}  \begin{threeparttable}
   \begin{tabular}{lcccccccc}
    \toprule \toprule
          & \multicolumn{2}{c}{{Coefficients}} &       & \multicolumn{2}{c}{{Variances}} &       & \multicolumn{2}{c}{{Correlations}} \\
 \cline{2-3}\cline{5-6}\cline{8-9}
          & \multicolumn{1}{c}{RMSE} & \multicolumn{1}{c}{MAE} &       & \multicolumn{1}{c}{RMSE} & \multicolumn{1}{c}{MAE} &       & \multicolumn{1}{c}{RMSE} & \multicolumn{1}{c}{MAE} \\
          \midrule
    MNP-FS  & 0.095  &  0.078  &&  0.231  &  0.170  & & 0.221   & 0.175 \\
    MNP-BN  & 0.132  &  0.111  &&  0.407  &  0.308  & & 0.275   & 0.212 \\
    MNP-MPR & 0.090  &  0.070  &&  0.481  &  0.357  & & 0.302   & 0.254 \\
   \bottomrule \bottomrule
    \end{tabular}%
\begin{tablenotes}
\footnotesize
\item This table reports the root mean squared error (RMSE) and mean absolute error (MAE) of the posterior mean estimates relative to the true parameter values. The panels for the coefficients, variances and correlations correspond to the error measures associated to the vector of coefficients $\beta_0$, the diagonal elements of $\Sigma_0$, and the implied correlations in $\Sigma_0$, respectively. The rows denote the alternative MNP model specifications considered.
\end{tablenotes}
\end{threeparttable}
  \label{tab:errormeasures}
\end{table}

\subsection{Sensitivity analysis base category specification}
\citet{burgette2019symmetric} show that the estimated probabilities from Bayesian MNP models can depend on the base category specification. This section analyses the sensitivity of the MNP-FS results to the base category specification under the identity prior $\theta=(0,${$1$}$,5)^\top$ and the equicorrelated prior $\theta=(${$1.525$}$,${$1$}$,5)^\top$ specifications and $q=1$.

\begin{figure}[tb!]
\caption{Purchase probabilities with different base categories}
\centering
\includegraphics[width=\textwidth]{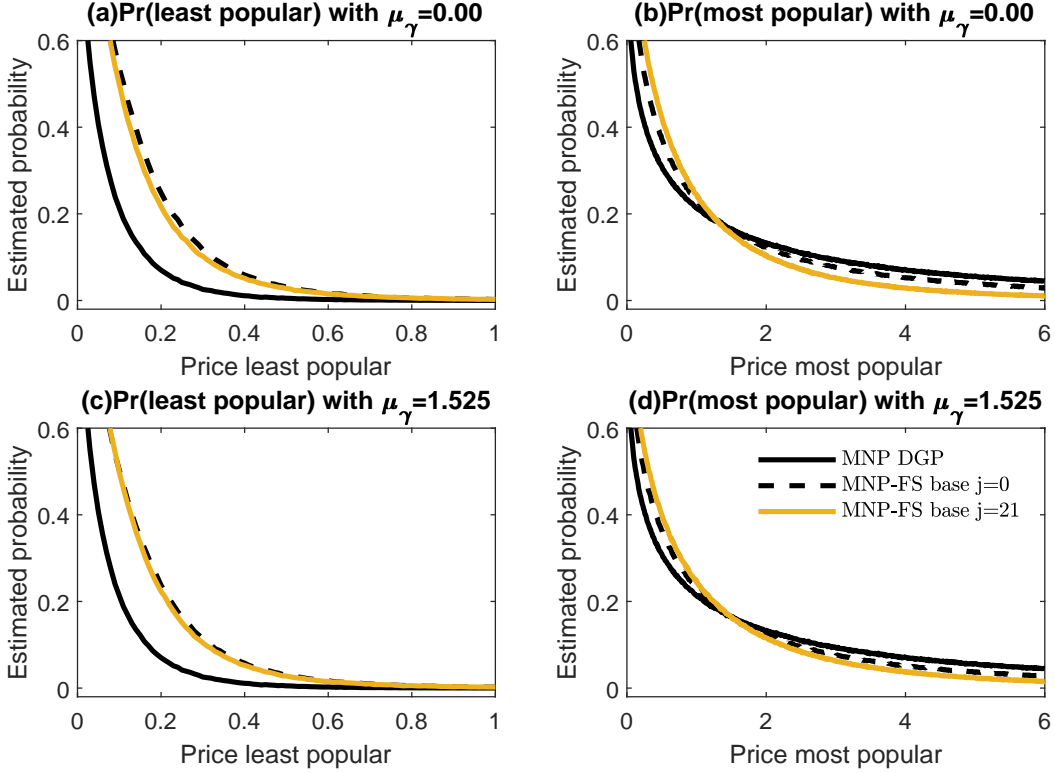}
 \fnote{This figure shows the estimated probabilities as a function of price, where panels (a) and (c) show the probabilities of the least popular category and panels (b) and (d) of the most popular category. The black line corresponds to the true probabilities. The estimated probabilities are based on the posterior parameter density of MNP-FS with $j=0$ as base category (dashed black line) and $j=21$ as base category (solid yellow line). Panels (a) and (b) show the results with prior specification $\theta=(0,${$1$}$,5)^\top$ and panel (c) and (d) with $\theta=(${$1.525$}$,${$1$}$,5)^\top$. }
\label{fig:simul_base}
\end{figure}

First, we consider the identity prior. Panel (a) in Figure~\ref{fig:simul_base} presents the estimated probabilities as a function of price for the least popular category, which has 16 observations. The solid black line corresponds to the true probabilities, the dashed black line is estimated using the same base category as in Section~\ref{sec:simulresults} which has 85 observations, while the yellow line corresponds to the specification that has the largest category with 440 observations as the base category.  In all panels, the price of the other categories is fixed at the mean across all observations.

Panel (a) in Figure~\ref{fig:simul_base} shows that the yellow line and the dashed black line are different from each other. Panel (b) presents the equivalent plot for the most popular category. In this panel, there is a slightly bigger difference between the yellow and dashed black line, indicating that the estimated probabilities of the least popular choice are less sensitive to the base category specification than the most popular category. 

As discussed in Section~\ref{sec: demonstration}, the identity prior does not necessarily imply an identity covariance matrix for the untransformed utilities. Since the utilities are in differences with the base category, this prior specification may result in estimates that are sensitive to the base category specification. In contrast, the equicorrelated prior follows from an identity covariance matrix for the untransformed utilities. 

Panels (c) and (d) show the estimated probabilities for the least and most popular categories, respectively, when using an equicorrelated prior. Panel (c) indicates that the results for the least popular category are not as sensitive as those when using an identity prior. The estimated probabilities from the different base category specifications are now almost identical. We find the same results for the most popular category in Panel (d). 

In sum, we conclude that the estimated probabilities are sensitive to the base category specification when a prior specification that shrinks $\Sigma$ to an identity matrix is employed. However, the impact of the base category can be substantially decreased by specifying an equicorrelated prior. An alternative way to deal with the sensitivity to the base category, is to pool the estimated probabilities across all models with different base category specifications to obtain probabilities that do not depend on one base category. This approach is computationally costly, especially when the number of choice alternatives is large.
\section{Empirical applications}\label{sec: application}
This section fits the multinomial probit model to consumer choice data sets of different dimensions. First, Section~\ref{sec:appsmall} considers a traditional consumer choice data set on laundry detergent purchases with only six alternatives. Section~\ref{sec:applarge} constructs a laundry detergent purchases data set with 50 choice alternatives from a big transaction data set. Section~\ref{sec:appmar} considers margarine brands in both a widely used small choice set and a newly constructed large choice set.

The proposed multinomial probit model is compared to benchmark specifications. These specifications and prior settings are discussed in Section~\ref{sec:design}. Moreover, we estimate a multinomial probit model with the covariance matrix fixed to the identity matrix, referred to as MNP-I.

The in-sample and out-of-sample predictive accuracy of the models are evaluated in terms of the predictive hit-rate and the logarithmic score (log-score). The predictive probability mass function for $Y_i$ is given by
\begin{align}\label{eq:predictive}
   p({Y}_i|X_i,Y,X,B,\theta) = \int p({Y}_i|X_i,\beta,\kappa)p(\beta,\kappa|Y,X,B,\theta)d\beta d\kappa,
\end{align}
where $X_i$ denotes the attributes of the observation $i$ to be predicted. For ease of notation we refer to $p({Y}_i|X_i,Y,X,B,\theta)$ as $p({Y}_i|X_i)$. An estimate $\hat{p}({Y}_i|X_i)$ for the predictive in \eqref{eq:predictive} is constructed as the empirical probability mass implied by the draws ${Y}_i^{[m]}$ obtained from $p(Y_i|X_i,\beta^{[m]},\kappa^{[m]})$, where  $\{\beta^{[m]}\}_{m=1}^M$ and $\{\kappa^{[m]}\}_{m=1}^M$ denote the MCMC draws.

The point forecast $\hat{Y}_i$ for $Y_i$ is constructed as the mode of $\hat{p}({Y}_i|X_i)$. The hit-rate is defined as
\begin{align}\label{eq:hitrate}
    \text{hit-rate} = \frac{1}{N}\sum_{i=1}^N I[\hat{Y}_i=Y_i],
\end{align}
where $I[A]$ is an indicator function. The log-score is defined as
\begin{align}\label{eq:logscore}
    \text{log-score} = \frac{1}{N}\sum_{i=1}^N&\ln(\hat{p}({Y}_i|X_i)).
\end{align}
For both the hit-rate and the log-score large values are preferred. As a rough measure of statistical significance, we test the difference of the hit-rates between MNP-FS and the benchmark models by a normal test and the difference of the log-scores by the \citet{giacomini2006tests} test. 

We randomly allocate 80\% of the observations for estimation of the model, and the remaining 20\% are employed for out-of-sample evaluation.

\subsection{Traditional data set with six choice alternatives}\label{sec:appsmall}
\citet{imai2005bayesian} and \citet{burgette2019symmetric} fit multinomial probit models to purchase data of laundry detergents. This data contains purchases of 2657 households out of six brands of laundry detergents, and the log price of each brand. The data set is described in detail by \citet{chintagunta1998empirical} and available in \citet{imai2005mnp}. We follow \citet{imai2005bayesian} and \citet{burgette2019symmetric} and fit the multinomial probit models with an intercept and the log price for each brand.

Table~\ref{tab:forecast} shows that for a small choice set the hit-rates and log-scores of our proposed model are similar to the ones of the benchmark methods. The symbol $(^+)$ indicates that MNP-FS significantly outperforms  the benchmark method, while the symbol $(^-)$ indicates that the benchmark significantly outperforms MNP-FS.
Only the in-sample log-score of MNP-BN significantly improves upon MNP-FS. All models show substantial improvements over the naive forecast, which is constructed as the observed in-sample frequency of the choice alternatives. However, MNP-FS, MNP-BN and MNP-MPR do not have a better in-sample hit-rate or out-of-sample log-score than the MNP-I.

\begin{table}[tb!]
  \centering \small
  \caption{Hit-rate and log-score for laundry detergent applications}
  \begin{threeparttable}
    \begin{tabular}{llrrrrr}
    \toprule \toprule
     \multicolumn{7}{c}{6 laundry detergent categories}\\
          \midrule
         Sample & Metric    & \multicolumn{1}{c}{ MNP-FS} & \multicolumn{1}{c}{MNP-BN} & \multicolumn{1}{c}{MNP-MPR} & \multicolumn{1}{c}{MNP-I} & \multicolumn{1}{c}{Naive} \\
          \midrule
    in    & hit-rate & 0.495 & 0.494\hspace{0.25cm} & 0.498 & 0.504\hspace{0.25cm} & 0.272$^{+}$\\
    in    & log-score & -1.328 & -1.325$^{-}$ & -1.326 & -1.349$^{+}$ & -1.641$^{+}$\\
    out   & hit-rate & 0.488 & 0.482\hspace{0.25cm} & 0.486 & 0.484\hspace{0.25cm} & 0.262$^{+}$ \\
    out   & log-score & -1.404 & -1.401\hspace{0.25cm} & -1.402 & -1.372\hspace{0.25cm} & -1.637$^{+}$ \\
              \midrule \midrule
          \multicolumn{7}{c}{50 laundry detergent categories}\\
          \midrule
           Sample & Metric    & \multicolumn{1}{c}{ MNP-FS} & \multicolumn{1}{c}{MNP-BN} & \multicolumn{1}{c}{MNP-MPR} & \multicolumn{1}{c}{MNP-I} & \multicolumn{1}{c}{Naive} \\
           \midrule
    in    & hit-rate & 0.219 & 0.226\hspace{0.25cm} & 0.213\hspace{0.25cm} & 0.165$^{+}$ & 0.058$^{+}$ \\
    in    & log-score & -3.301 & -3.306\hspace{0.25cm} & -3.409$^{+}$ & -3.725$^{+}$ & -3.792$^{+}$ \\
    out   & hit-rate & 0.209 & 0.206\hspace{0.25cm} & 0.204\hspace{0.25cm} & 0.154$^{+}$ & 0.055$^{+}$ \\
    out   & log-score & -3.339 & -3.350$^{+}$ & -3.456$^{+}$ & -3.752$^{+}$ & -3.761$^{+}$ \\
    \bottomrule \bottomrule
    \end{tabular}%
\begin{tablenotes}
\footnotesize
\item This table shows the in- and out-of-sample hit-rates and log-scores, defined in respectively \eqref{eq:hitrate} and \eqref{eq:logscore}. Predictive densities are estimated on the data discussed in Section~\ref{sec:appsmall} and Section~\ref{sec:applarge} using the multinomial probit model with a trace-restricted factor structure (MNP-FS), with a trace-restriction (MNP-BN), with $\Sigma_{11}=1$ (MNP-MPR), with $\Sigma=I_J$ (MNP-I), and a naive method in which the forecast equals the most frequently observed category in the data. The symbol ($^{+}$) indicates that MNP-FS performs significantly better than the method indicated by the column label and the symbol ($^{-}$) that MNP-FS performs significantly worse, on a significance level of 5\%. Appendix~\ref{sec:addapp} {reports} the {corresponding} p-values of the pairwise tests.
\end{tablenotes}
\end{threeparttable}
  \label{tab:forecast}
\end{table}

\begin{figure}[tb!]
\caption{Posterior mean of the correlation matrix with six choice alternatives}
\centering
\includegraphics*[width=\textwidth,trim = 0 0 0 0]{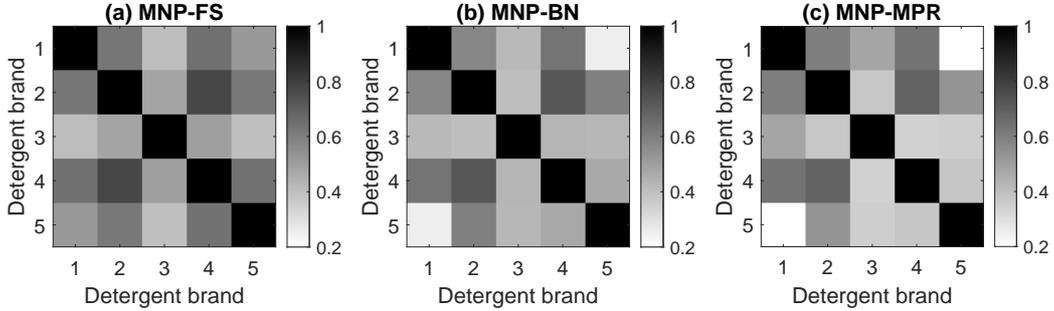}
 \fnote{This figure shows the posterior mean estimates of the elements of the correlation matrix of the latent utilities of six laundry detergent brands. Panel (a) shows the posterior mean of the multinomial probit model with a trace-restricted factor structure (MNP-FS), (b) with a trace-restriction (MNP-BN), and (c) with $\Sigma_{11}=1$ (MNP-MPR).
 }
\label{fig:cormat6}
\end{figure}

\begin{figure}[tb!]
\caption{Posterior density of the price coefficient with six choice alternatives}
\centering
\includegraphics*[width=\textwidth,trim = 0 0 0 0]{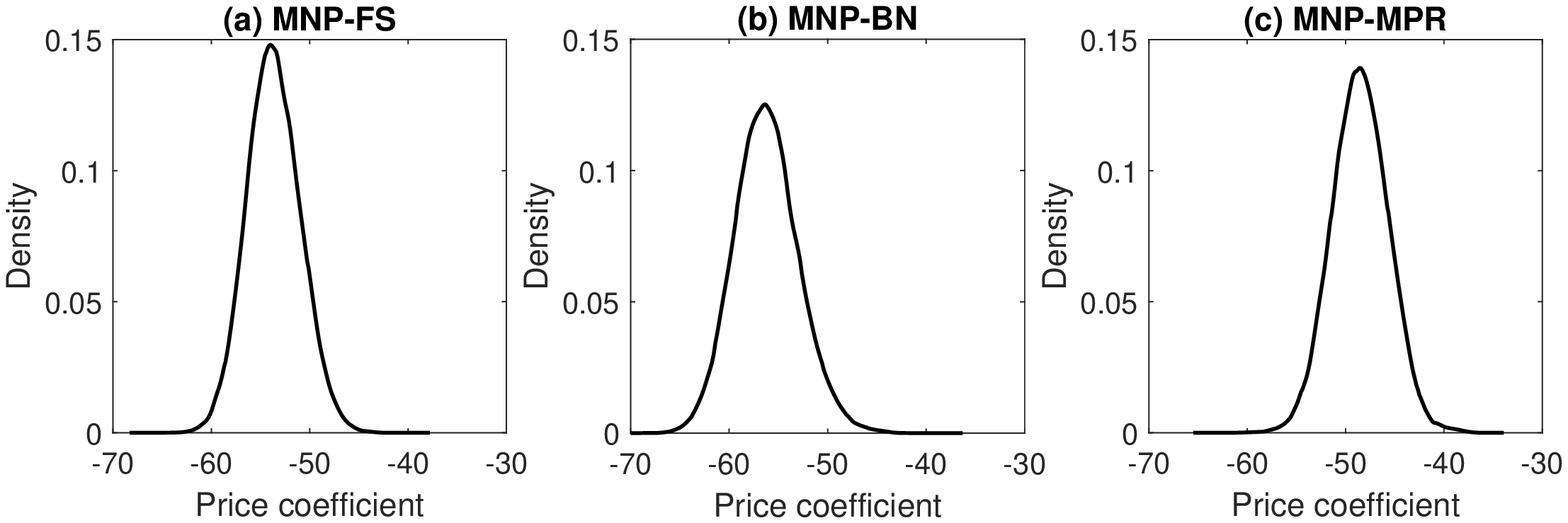}
 \fnote{This figure shows the posterior densities of the coefficient of the log prices of six laundry detergent brands. Panel (a) shows the posterior density of the multinomial probit model with a trace-restricted factor structure (MNP-FS), Panel (b) with a trace-restriction (MNP-BN), and Panel (c) with $\Sigma_{11}=1$ (MNP-MPR).
 }
\label{fig:price6}
\end{figure}

The posterior parameter estimates from MNP-FS, MNP-BN and MNP-MPR are also similar. For instance, the posterior mean of the correlation matrix of the latent utilities presented in Figure~\ref{fig:cormat6}, has similar patterns across all three methods. The small differences in the posterior estimates  are also observed for the price coefficient. Figure~\ref{fig:price6} shows the posterior densities for the different models. The densities are concentrated around similar values.

\subsection{Modern data set with 50 choice alternatives}\label{sec:applarge}
Nowadays, almost all real-life consumer choice sets contain many more choice alternatives than six. To illustrate the importance of a scalable multinomial probit model in these settings, we analyse a laundry detergent purchase data set with 50 choice alternatives.

We use the Complete Journey dataset published by Dunnhumby\footnote{https://www.dunnhumby.com/sourcefiles}. This dataset contains all purchases of 92,339 products over two years from a group of 2,500 households at a retailer. We filter the purchases of products with the description ``Laundry Detergents", which results in 300 unique products with different brands, sizes and variants such as liquid or powder detergents. Since the unique products with a small purchase volume are of less interest to a marketing manager, we focus on the 50 top-selling products. 

We define the log price of each brand in the same way as, for instance, \citet{allenby1991quality} and \citet{wan2017modeling}. The Dunnhumby data set only contains records of shelf prices at purchase dates. We impute the prices for products that are not sold on a certain purchase date by taking the mean of the observed prices of a specific product on the nearest date in the same week. In {cases where} there is no purchase record in the same week, we take the most recent observed price. We remove the observations for which we cannot impute a price for each product. 

The final sample contains 4839 observations on 50 categories, which contains 64\% of the laundry detergent purchases and the purchase frequency varies from 30 to 274 per category. We fit the same models as in the exercise with six choice alternatives, also considering an intercept and log price coefficient.

For this large choice set, Table~\ref{tab:forecast} shows that the in-sample and out-of-sample log-score of our proposed model are larger than those of the benchmark models, with the out-of-sample improvement also being statistically significant. The hit-rates are not statistically different from those of MNP-BN and MNP-MPR. However, the MNP-FS reports both hit-rates and log-scores significantly larger than that of MNP-I. Comparing this result to the relative performance of MNP-FS and MNP-I with six choice alternatives, suggests that accounting for correlations across utilities is especially important when the choice set is large. 

\begin{figure}[tb!]
\caption{Posterior mean of the correlation matrix with 50 choice alternatives}
\centering
\includegraphics*[width=\textwidth,trim = 0 0 0 0]{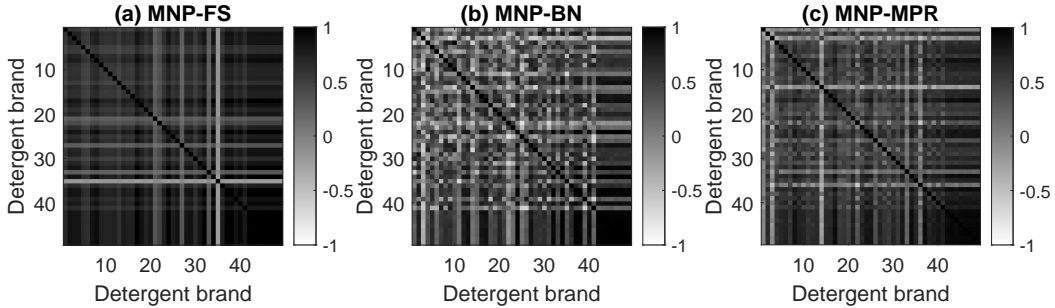}
 \fnote{This figure shows the posterior means of the elements of the correlation matrix $\Sigma$ of the latent utilities of 50 unique laundry detergent products. Panel (a) shows the posterior mean of the multinomial probit model with a trace-restricted factor structure (MNP-FS), (b) with a trace-restriction (MNP-BN), and (c) with $\Sigma_{11}=1$ (MNP-MPR).
 }
\label{fig:cormat50}
\end{figure}

For the large laundry detergent choice set, the posterior parameter estimates show differences across the different models. Figure~\ref{fig:cormat50} presents the posterior means of the elements of the correlation matrix of the latent utilities. While some general patterns are common to all three methods, the posterior correlations of MNP-BN and MNP-MPR show more variation. The factor structure in MNP-FS has fewer parameters, which restrict the patterns in the correlation matrix. Parsimony may become more important in this large choice set, as the number of estimated covariance parameters by MNP-BN and MNP-MPR equals $49\times50/2=1225$ relative to 4839 observations. This might explain the fact that the predictive accuracy of the MNP-FS, which only estimates $49\times2=98$ covariance parameters, significantly improves upon benchmark models in the large choice set, while the differences are not significant in the small choice set.

\begin{figure}[tb!]
\caption{Posterior density of the price coefficient with 50 choice alternatives}
\centering
\includegraphics*[width=\textwidth,trim = 0 0 0 0]{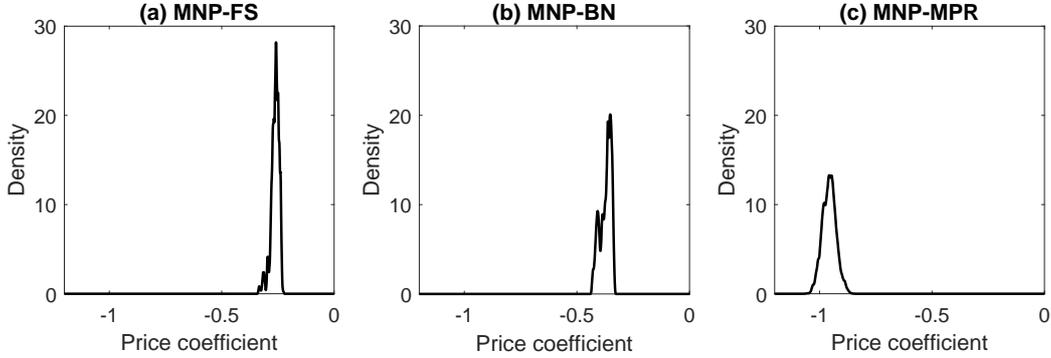}
 \fnote{This figure shows the posterior densities of the coefficient of the log prices of 50 unique laundry detergent products. Panel (a) shows the posterior density of the multinomial probit model with a trace-restricted factor structure (MNP-FS), (b) with a trace-restriction (MNP-BN), and (c) with $\Sigma_{11}=1$ (MNP-MPR).
 }
\label{fig:price50}
\end{figure}

Figure~\ref{fig:price50} reports the posterior densities for the price coefficient. We find that the differences in the specifications of the covariance matrix are also reflected in the posterior of the price coefficient. The posterior mean of the price coefficient in the MNP-FS model equals -0.262, compared to -0.372 and -0.958 in the MNP-BN and MNP-MPR respectively. The corresponding posterior standard deviation is respectively 0.018, 0.025, and 0.031. Hence we conclude that the benchmark specifications result in a larger price effect estimate with more posterior uncertainty than the MNP-FS model.

\begin{figure}[tb!]
\caption{Purchase probabilities estimated with 6 and 50 choice alternatives}
\centering
\includegraphics*[width=\textwidth,trim = 0 0 0 0]{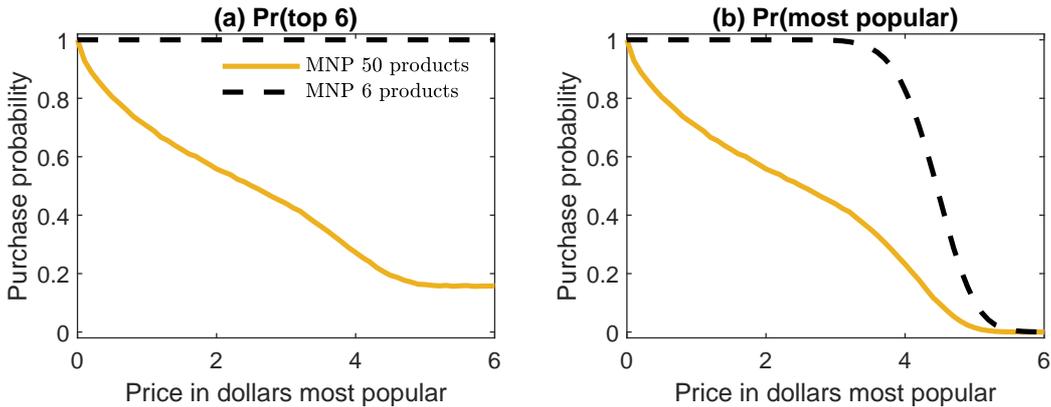}\\[3mm]
 \fnote{This figure shows the purchase probabilities as a function of price estimated using all 50 products (yellow solid line) and only the top six selling products (black dashed line). Panel (a) shows the probability of buying one of the top six selling products, and Panel (b) the probability of buying the most popular product conditional on buying a top six product. The probabilities vary with the price of the most popular product. All probabilities are estimated in the multinomial probit model with a trace-restricted factor structure.
 }
\label{fig:price50to6}
\end{figure}

One might argue that in most practical settings only a small set of high volume products are of interest. To examine the importance of considering a large choice set, we compare the effect of price on the purchase probability of the six most popular products between an MNP model estimated on only those six products and an MNP model estimated on the total choice set. Figure~\ref{fig:price50to6} suggest that, even when only the six top selling products are of interest, including the other products in the model is important for an effective pricing strategy.

Panel (a) of Figure~\ref{fig:price50to6} shows the probability of buying one of the top six selling products as a function of the price of the most popular product. The dashed black line corresponds to MNP-FS with only the top six products included as choice alternatives, and the solid yellow line to MNP-FS with 50 products included. The model that includes all 50 products indicates that the probability of buying in the top six decreases when the price of the top product increases. In other words, consumers substitute away from the top product to products outside the top six. This effect cannot be captured by the model that only includes the top six products, which sets the probability of buying in the top six equal to one by construction.

Panel (b) of Figure~\ref{fig:price50to6} shows the probability of buying the most popular product as a function of its price. The dashed black line shows that demand for laundry detergent is inelastic between a price of zero and four, and highly elastic for prices higher than four.
However, the solid yellow line shows that including the purchases of all 50 products, results in purchase probabilities that change smoothly with price. This result suggests that excluding purchases of low volume products from the analysis can bias the estimated price effect of high volume products.

\subsection{Margarine purchases}\label{sec:appmar}
We compare the hit-rates and log-scores of the MNP-FS to the benchmark methods on another commonly used choice data set. \citet{mcculloch1994exact}, \citet{burgette2012trace}, and \citet{burgette2019symmetric} fit multinomial probit models to panel data on purchases of margarine by 516 households. They estimate an intercept and a log price coefficient for six brands. The data is described in detail by \citet{allenby1991quality} and available in \citet{rossi2012bayesian}. We follow \citet{burgette2012trace} and \citet{burgette2019symmetric} and fit the multinomial probit models to the first purchase of each household.

As an alternative for the small margarine choice set, we construct a large margarine choice data set in the same way as for laundry detergents in Section~\ref{sec:applarge}. We filter the purchases of products with the description ``Margarines" from the Complete Journey dataset which results in 178 unique products. The final sample contains 11754 observations on 50 categories, which contains 96\% of the margarine purchases and the purchase frequency varies from 32 to 1206 per category.

Table~\ref{tab:forecastmar} shows that for the margarine data our proposed model has higher in-sample and out-of-sample hit-rates and log-scores than the benchmark models when the number of choice alternatives is large. The log-scores of the MNP-FS are significantly larger than the log-scores of all the benchmark models. The MNP-FS model does not show significantly different performance from the MNP-I in the small choice set, but shows significant improvements in the large choice set on all metrics. This result is in line with the laundry detergent application and supports the claim that it is important to take correlations into account in large choice sets.

\begin{table}[tb!]
  \centering \small
  \caption{Hit-rate and log-score for margarine applications}
  \begin{threeparttable}
    \begin{tabular}{llrrrrr}
    \toprule \toprule
          \multicolumn{7}{c}{6 margarine categories}\\
          \midrule
           Sample & Metric    & \multicolumn{1}{c}{ MNP-FS} & \multicolumn{1}{c}{MNP-BN} & \multicolumn{1}{c}{MNP-MPR} & \multicolumn{1}{c}{MNP-I} & \multicolumn{1}{c}{Naive} \\
          \midrule
in    & hit-rate & 0.466 & 0.466\hspace{0.25cm} & 0.468 & 0.478 & 0.426 \hspace{0.1cm} \\
    in    & log-score & -1.450 & -1.448\hspace{0.25cm} & -1.452 & -1.453 & -1.581$^{+}$ \\
    out   & hit-rate & 0.624 & 0.624\hspace{0.25cm} & 0.624 & 0.634 & 0.584 \hspace{0.1cm} \\
    out   & log-score & -1.269 & -1.264$^{-}$ & -1.270 & -1.270 & -1.323 \hspace{0.1cm} \\
              \midrule \midrule
          \multicolumn{7}{c}{50 margarine categories}\\
          \midrule
           Sample & Metric    & \multicolumn{1}{c}{ MNP-FS} & \multicolumn{1}{c}{MNP-BN} & \multicolumn{1}{c}{MNP-MPR} & \multicolumn{1}{c}{MNP-I} & \multicolumn{1}{c}{Naive} \\
           \midrule
    in    & hit-rate & 0.358 & 0.345\hspace{0.25cm} & 0.335$^{+}$ & 0.232$^{+}$ & 0.106$^{+}$ \\
    in    & log-score & -2.817 & -2.826$^{+}$ & -2.894$^{+}$ & -3.462$^{+}$ & -3.547$^{+}$ \\
    out   & hit-rate & 0.341 & 0.331 \hspace{0.1cm} & 0.323 \hspace{0.1cm} & 0.209$^{+}$ & 0.095$^{+}$ \\
    out   & log-score & -2.850 & -2.863$^{+}$ & -2.939$^{+}$ & -3.505$^{+}$ & -3.561$^{+}$ \\
    \bottomrule \bottomrule
    \end{tabular}%
\begin{tablenotes}
\footnotesize
\item This table shows the in- and out-of-sample hit-rates and log-scores for the margarine data sets discussed in Section~\ref{sec:appmar}. See Table~\ref{tab:forecast} for details. 
\end{tablenotes}
\end{threeparttable}
  \label{tab:forecastmar}
\end{table}

Table~\ref{tab:forecastmar_application} in Appendix~\ref{sec:addapp} reports the hit-rates and log-scores for MNP-FS when the equicorrelated prior is considered. The results are similar and indicate that employing an equicorrelated prior does not necessarily lead to an increase in predictive performance.











\section{Conclusion}\label{sec: conclusion}
This paper proposes a factor structure on the covariance matrix in the multinomial probit model that makes the model scalable to the dimensions of modern choice sets. The model parameters are identified by a reparamatrization of the factor structure that imposes a trace-restriction on  the covariance matrix. 

A numerical experiment shows that the model parameters can be accurately estimated on a choice set with 50 alternatives in the proposed multinomial probit specification. On a real data set with 50 choice alternatives, the hit-rates and log-scores demonstrate {significant predictive} improvements relative to benchmark approaches.

The large size of modern assortments and the increasing amount of product differentiation makes the scalable choice model that accounts for correlations across choice alternatives of managerial relevance. The empirical application to retail data of 50 laundry detergents suggests that managers may overestimate the price effect when only analysing top-selling products, relative to including all products in the multinomial probit model. 


\bibliographystyle{apalike} 
\bibliography{mnp}

\begin{thebibliography}{}

\bibitem[Allenby and Rossi, 1991]{allenby1991quality}
Allenby, G.~M. and Rossi, P.~E. (1991).
\newblock Quality perceptions and asymmetric switching between brands.
\newblock {\em Marketing science}, 10(3):185--204.

\bibitem[Bunch, 1991]{bunch1991estimability}
Bunch, D.~S. (1991).
\newblock Estimability in the multinomial probit model.
\newblock {\em Transportation Research Part B: Methodological}, 25(1):1--12.

\bibitem[Burgette and Nordheim, 2012]{burgette2012trace}
Burgette, L.~F. and Nordheim, E.~V. (2012).
\newblock The trace restriction: An alternative identification strategy for the
  {B}ayesian multinomial probit model.
\newblock {\em Journal of Business \& Economic Statistics}, 30(3):404--410.

\bibitem[Burgette et~al., 2021]{burgette2019symmetric}
Burgette, L.~F., Puelz, D., and Hahn, P.~R. (2021).
\newblock A symmetric prior for multinomial probit models.
\newblock {\em Bayesian Analysis}, pages 1--18.

\bibitem[Burgette and Reiter, 2013]{burgette2013multiple}
Burgette, L.~F. and Reiter, J.~P. (2013).
\newblock Multiple-shrinkage multinomial probit models with applications to
  simulating geographies in public use data.
\newblock {\em Bayesian Analysis}, 8(2):453--478.

\bibitem[Chintagunta and Prasad, 1998]{chintagunta1998empirical}
Chintagunta, P.~K. and Prasad, A.~R. (1998).
\newblock An empirical investigation of the “dynamic {McFadden}” model of
  purchase timing and brand choice: Implications for market structure.
\newblock {\em Journal of Business \& Economic Statistics}, 16(1):2--12.

\bibitem[Cripps et~al., 2009]{cripps2009parsimonious}
Cripps, E., Fiebig, D.~G., and Kohn, R. (2009).
\newblock Parsimonious estimation of the covariance matrix in multinomial
  probit models.
\newblock {\em Econometric Reviews}, 29(2):146--157.

\bibitem[Geweke et~al., 1994]{geweke1994alternative}
Geweke, J., Keane, M., and Runkle, D. (1994).
\newblock Alternative computational approaches to inference in the multinomial
  probit model.
\newblock {\em The review of economics and statistics}, pages 609--632.

\bibitem[Geweke and Zhou, 1996]{geweke1996measuring}
Geweke, J. and Zhou, G. (1996).
\newblock Measuring the pricing error of the arbitrage pricing theory.
\newblock {\em The review of financial studies}, 9(2):557--587.

\bibitem[Giacomini and White, 2006]{giacomini2006tests}
Giacomini, R. and White, H. (2006).
\newblock Tests of conditional predictive ability.
\newblock {\em Econometrica}, 74(6):1545--1578.

\bibitem[Hausman and McFadden, 1984]{hausman1984specification}
Hausman, J. and McFadden, D. (1984).
\newblock Specification tests for the multinomial logit model.
\newblock {\em Econometrica: Journal of the Econometric Society}, pages
  1219--1240.

\bibitem[Imai and Van~Dyk, 2005a]{imai2005bayesian}
Imai, K. and Van~Dyk, D.~A. (2005a).
\newblock A {B}ayesian analysis of the multinomial probit model using marginal
  data augmentation.
\newblock {\em Journal of Econometrics}, 124(2):311--334.

\bibitem[Imai and Van~Dyk, 2005b]{imai2005mnp}
Imai, K. and Van~Dyk, D.~A. (2005b).
\newblock {MNP}: R package for fitting the multinomial probit model.
\newblock {\em Journal of Statistical Software}, 14(3):1--32.

\bibitem[Lopes, 2014]{lopes2014modern}
Lopes, H.~F. (2014).
\newblock Modern {B}ayesian factor analysis.
\newblock {\em Bayesian Inference in the Social Sciences}, pages 115--153.

\bibitem[McCulloch and Rossi, 1994]{mcculloch1994exact}
McCulloch, R. and Rossi, P.~E. (1994).
\newblock An exact likelihood analysis of the multinomial probit model.
\newblock {\em Journal of Econometrics}, 64(1-2):207--240.

\bibitem[McCulloch et~al., 2000]{mcculloch2000bayesian}
McCulloch, R.~E., Polson, N.~G., and Rossi, P.~E. (2000).
\newblock A {B}ayesian analysis of the multinomial probit model with fully
  identified parameters.
\newblock {\em Journal of {E}conometrics}, 99(1):173--193.

\bibitem[Piatek and Gensowski, 2017]{piatek2017multinomial}
Piatek, R. and Gensowski, M. (2017).
\newblock A multinomial probit model with latent factors: Identification and
  interpretation without a measurement system.
\newblock {\em IZA Discussion Paper}.

\bibitem[Roberts and Rosenthal, 2009]{roberts2009examples}
Roberts, G.~O. and Rosenthal, J.~S. (2009).
\newblock Examples of adaptive {MCMC}.
\newblock {\em Journal of Computational and Graphical Statistics},
  18(2):349--367.

\bibitem[Rossi et~al., 2012]{rossi2012bayesian}
Rossi, P.~E., Allenby, G.~M., and McCulloch, R. (2012).
\newblock {\em Bayesian statistics and marketing}.
\newblock John Wiley \& Sons.

\bibitem[Smith, 2015]{smith2015copula}
Smith, M.~S. (2015).
\newblock Copula modelling of dependence in multivariate time series.
\newblock {\em International Journal of Forecasting}, 31(3):815--833.

\bibitem[Wan et~al., 2017]{wan2017modeling}
Wan, M., Wang, D., Goldman, M., Taddy, M., Rao, J., Liu, J., Lymberopoulos, D.,
  and McAuley, J. (2017).
\newblock Modeling consumer preferences and price sensitivities from
  large-scale grocery shopping transaction logs.
\newblock In {\em Proceedings of the 26th International Conference on World
  Wide Web}, pages 1103--1112.

\bibitem[Yeo and Johnson, 2000]{yeo2000new}
Yeo, I.-K. and Johnson, R.~A. (2000).
\newblock A new family of power transformations to improve normality or
  symmetry.
\newblock {\em Biometrika}, 87(4):954--959.

\end{thebibliography}
\clearpage
\appendix
\section{Approximating prior distribution}\label{App:FlexibleDistribution}
The density $\tilde{p}(\kappa_l|\lambda_l)$ is constructed through a transformation of $\kappa_l$. Let $t_{\eta_l}: \mathbb{R}\rightarrow \mathbb{R}$ be a differentiable monotonic function with parameter vector $\eta_l$. Let $G: \Omega_l\rightarrow\mathbb{R}$ be a differentiable monotonic function mapping from the support of $\kappa_l$, $\Omega_l$, to the real line.
Now, consider the transformation $x_l = t_{\eta_l}\left[\frac{1}{\tau_l}\left(G(\kappa_l)-\mu_l\right)\right]$, where  $\tau_l>0$ and $\mu_l$ are scalars, and $x_l\sim N(0,1)$.  The implied distribution on $\kappa_l$ can be recovered by the Jacobian of the transformation from $x_l$ to $\kappa_l$ so that
\begin{equation}
\tilde{p}(\kappa_l|\lambda_l) = \phi_1\left\{t_{\eta_l}\left[\frac{G\left(\kappa_l\right)-\mu_l}{\tau_l}\right]\right\}t_{\eta_l}'\left[\frac{G\left(\kappa_l\right)-\mu_l}{\tau_l}\right]\frac{1}{\tau_l}G'\left(\kappa_l\right),
\end{equation}
where $\lambda_l = \left(\mu_l,\tau_l,\eta_l\right)^\top$, $\phi_1$ denotes the density function of a standard normal distribution, while $t_{\eta_l}'(.)$ and $G'(.)$ denote the first derivative of $t_{\eta_l}$ and $G$, respectively.
The role of $G$ is to transform $\kappa_l$ into the real line.
We employ $G\left(\kappa_l\right) = \Phi_1^{-1}\left(\frac{\kappa_l}{\pi}\right)$ for $l<n-1$, and $G\left(\kappa_l\right) = \Phi_1^{-1}\left(\frac{\kappa_l}{2\pi}\right)$  for $l=n-1$.
The role of transformation $t_{\eta_i}$ is to induce a family of density functions, $\tilde{p}(\kappa_l|\lambda_l)$, capable of accurately approximating the prior $p(\kappa_l|\theta)$. With this goal in mind, $t_{\eta_l}$ is chosen to be the transformation suggested by \cite{yeo2000new}, proven effective to transform into near normality, as it is required here for $x_l\sim N(0,1)$. This transformation is defined as
 \[
 t_{\eta}(\nu)=
 \left\{\begin{array}{cl}
 -\frac{(-\nu+1)^{2-{\eta}}-1}{2-{\eta}} &\mbox{if }\nu<0,\\
 \frac{(\nu+1)^{\eta} -1}{\eta} &\mbox{if }\nu\geq 0,\,
 \end{array} \right.
 \]
 and its first derivative is computed as
 \[
 t_{\eta}'(\nu)=
 \left\{\begin{array}{cl}
 (-\nu+1)^{1-{\eta}} &\mbox{if }\nu<0,\\
 (\nu+1)^{{\eta}-1} &\mbox{if }\nu\geq 0.\,
 \end{array} \right.
 \]
 
     \section{Constructing the equicorrelated prior}\label{sec:equicorr}
For $i\ne j$ denote the prior mean of the correlation $\rho_{i,j}$ as
$$E_\theta\left(\rho_{i,j}\right) = E_\theta\left[\frac{\Sigma_{i,j}}{\sqrt{\Sigma_{i,i}\Sigma_{j,j}}}\right] = E_{\mu_\gamma,\sigma_\gamma,\nu}\left[\rho_{i,j}\right]$$
where $E_\theta$ is an expectation computed with respect to $p\left(\Sigma|\theta\right)$. The objective is to find the value of $\mu_\gamma$ for which $E_{\mu_\gamma,\sigma_\gamma,\nu}\left(\rho_{i,j}\right) = \frac{1}{2}$. For any fixed values of $\sigma_\gamma$, $\nu$ and $q$, this value can be found as the solution to the optimization problem
$$\mu_\gamma^* = \argmin_{\mu_\gamma\in\mathbb{R}^{+}} \left|\frac{1}{2}-E_{\mu_\gamma,\sigma_\gamma,\nu}\left(\rho_{i,j}\right)\right|.$$
Solution to this problem requires evaluation of the expectation $E_{\mu_\gamma,\sigma_\gamma,\nu}\left(\rho_{i,j}\right)$. We evaluate this expectation in a Monte Carlo fashion, by generating $100$ thousand draws from the prior and then computing the sample mean. To solve the optimization problem we use an off the shelf root finding algorithm.

\section{Details on the sampling scheme}\label{sec:mcmc}
In this appendix we discuss the steps of the MCMC sampling scheme in more detail.
To initialize $\kappa$ we use a draw from its prior distribution. The latent utilities $Z_i$ for $i=1,\dots,N$, are initialised by first sampling a standard normally-distributed vector $\tilde{Z}_i$ of length $J+1$ and center it at zero. The elements of $\tilde{Z}_i$ are then permuted until the largest element of $\tilde{Z}_i$ is located in row $y_i+1$. The initial latent utilities are set as $z_{ij}=\tilde{z}_{ij+1}-\tilde{z}_{i1}$ for $j=1,\dots,J$. Once all the parameters are initialised, we iterate over the following three steps.\\
 \ \\
\ \ \hspace{2cm} \textbf{Step 1}: Generate from $\beta|Z,\Sigma(\kappa),X$.  \\ 
Sampling of the coefficients $\beta$ is performed using the standard Gibbs sampling steps (see for instance \citealt{mcculloch1994exact}). Specifically, $\beta$ is generated from
    \begin{align}
        \beta|Z,\Sigma(\kappa),X\sim\mathcal{N}(\bar{b},\bar{B}^{-1}),
    \end{align}
   with $\bar{B}={X^*}^\top{X^*}+B$ and $\bar{b}=\bar{B}^{-1}{X^*}^\top{Z^*}$, where $X^*=(X_1^\top C,\dots,X_N^\top C)^\top$ and $Z^*=(Z_1^\top C,\dots,Z_N^\top C)^\top$, with $\Sigma^{-1}=CC^\top$.\\
   \ \\
\ \ \hspace{2cm} \textbf{Step 2}: Generate from $Z|\beta,\Sigma(\kappa),Y,X$.\\
To generate from the latent utilities we employ the truncated normal distributions as in \citet{mcculloch1994exact}.
    \begin{align}
        z_{ij}&\sim \mathcal{N}^+_{\max(Z_i^{(j)},0)}(X_{ij}^\top\beta + F(Z_i^{(j)}-X_{i(j)}\beta),\Sigma_{jj}-F\Sigma_{(j)j}),  \text{ if }  Y_i=j,\\
        z_{ij}&\sim\mathcal{N}^-_{\max(Z_i^{(j)},0)}(X_{ij}^\top\beta + F(z_i^{(j)}-X_{i(j)}\beta),\Sigma_{jj}-F\Sigma_{(j)j}),  \text{ if }  Y_i\neq j,
    \end{align}
    with $Z_i^{(j)}=(z_{i1},\dots,z_{ij-1},z_{ij+1},\dots,z_{iJ})$, and $\mathcal{N}^+_{a}(\mu,\sigma^2)$ and $\mathcal{N}^-_{a}(\mu,\sigma^2)$ represent a normal distribution with mean $\mu$ and variance $\sigma^2$ truncated from below or above by $a$, respectively. Here, $X_{ij}$ denotes the $j$ element in $X_i$, $X_{i(j)}$ denotes $X_i$ after removing $X_{ij}$. On the other hand, $F=\Sigma_{j(j)}\Sigma_{(j)(j)}^{-1}$, where  $\Sigma_{j(j)}$ denotes the $j$ row vector of $\Sigma$ without element $j$, and $\Sigma_{(j)(j)}$ denotes $\Sigma$ after removing row and column $j$.\\
 \ \\
\ \ \hspace{2cm} \textbf{Step 3}: Generate from $\kappa|Z,\beta,X$\\
Sampling of the parameters $\kappa$ is obtained via blocked random walk Metropolis-Hastings steps. 
At the start of each iteration, allocate the elements of $\kappa$ into $G$ parameter blocks, $\kappa_{b_1},\dots,\kappa_{b_G}$, of five elements each. For $g = 1,\dots,G$, generate a draw $\kappa_{b_g}^{\text{new}}$ from the proposal density,
    $$q(\kappa_{b_g}|\kappa_{b_g}^{\text{old}}) = \prod_{l=1}^5\frac{\phi_1\left(\kappa_{b_{g_l}};\kappa_{b_{g_l}}^{\text{old}},\sigma_{g_l}^2\right)}{\Phi_1\left(\text{up}_{g_l};\kappa_{b_{g_l}}^{\text{old}},\sigma_{g_l}^2\right)-\Phi_1\left(\text{low}_{g_l};\kappa_{b_{g_l}}^{\text{old}},\sigma_{g_l}^2\right)}.$$
    Accept $\kappa_{b_g}^{\text{new}}$  with probability
    \begin{align}
        \alpha = \min\left(1,\frac{p(\kappa_{b_g}^{\text{new}}|Z,\beta,X,\left\{\kappa\backslash\kappa_{b_g}\right\})q(\kappa_{b_g}^{\text{old}}|\kappa_{b_g}^{\text{new}})}{p(\kappa_{b_g}^{\text{old}}|Z,\beta,X,\left\{\kappa\backslash\kappa_{b_g}\right\})q(\kappa_{b_g}^{\text{new}}|\kappa_{b_g}^{\text{old}})}\right),
    \end{align}
    where
    \begin{align}
        p(\kappa_{b_g}|Z,\beta,X,\left\{\kappa\backslash\kappa_{b_g}\right\})\propto p(\kappa_{b_g}|\hat{\lambda})p(Z|X,\beta,\Sigma(\kappa)).
    \end{align}
    Here, $p(\kappa_{b_g}|\hat{\lambda})$ denotes the prior density for $\kappa_{b_g}$, while $p(Z|X,\beta ,\Sigma(\kappa))$ is the density of a multivariate normal distribution with mean $X\beta $ and covariance matrix  $\Sigma(\kappa) = \gamma(\kappa)\gamma(\kappa)^\top+D(\kappa)^2$.
    The expression $\left\{\kappa\backslash\kappa_{b_g}\right\}$ denotes the subtraction of the subset $\kappa_{b_g}$ from $\kappa$. The constants $\text{low}_{l}$ and $\text{up}_{l}$ denote the lower and upper bounds of $\kappa_l$. The proposal parameters $\sigma_l^2$ are set adaptively to target acceptance rates between $15\%$ and $30\%$. The random allocation of $\kappa$ into groups, plus the parameter-specific adaptive steps, allow the blocked sampler to target parameter-specific acceptance rates.
    
\section{Additional results: numerical experiment}\label{sec:addnum}
\begin{figure}[H]
\caption{Posterior mean parameters in numerical experiment}
\centering
\includegraphics*[width=\textwidth,trim = 0 0 0 0]{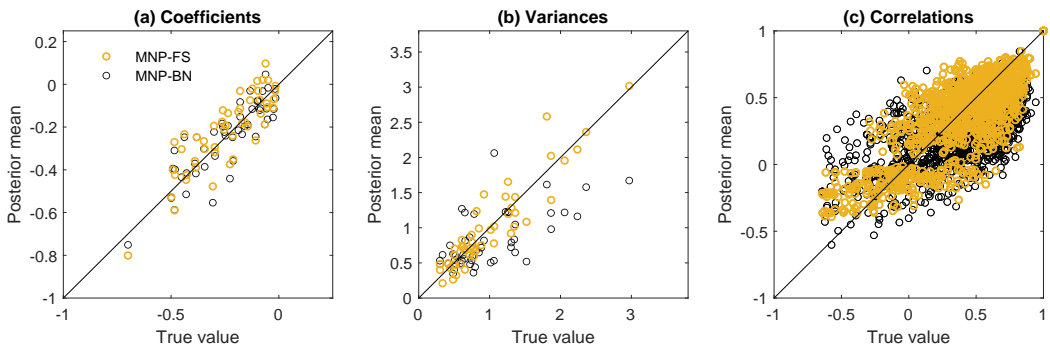}
 \fnote{{This figure presents the estimated posterior means from the MNP-FS} (yellow circles) and {MNP-MPR}(black circles) approaches, for the coefficients $\beta$ in Panel (a), and the variances and correlations of the latent utilities in $\Sigma$ in Panel (b) and (c), respectively.}
\label{fig:simulMNPR}
\end{figure}

\section{Additional results empirical applications}\label{sec:addapp}
\begin{table}[h!]
	\centering \small
	\caption{P-values of the tests of predictive accuracy}
	\begin{threeparttable}
		\begin{tabular}{llrrrr}
			\toprule\toprule
			                                                        \multicolumn{6}{c}{6 laundry detergent categories}                                                          \\ \midrule
			Sample & Metric    &  \multicolumn{1}{c}{MNP-BN} & \multicolumn{1}{c}{MNP-MPR} & \multicolumn{1}{c}{MNP-I} & \multicolumn{1}{c}{Naive} \\ \midrule
			in     & hit-rate  &                     0.951 &   0.854  &  0.560  &  0.000 \\
			in     & log-score &                     0.001 &   0.126  &  0.000  &  0.000 \\
			out    & hit-rate  &                     0.854 &   0.951  &  0.902  &  0.000 \\
			out    & log-score &                     0.057 &   0.511  &  0.512  &  0.002 \\ \midrule\midrule
			                           \multicolumn{6}{c}{50 laundry detergent categories}                                                         \\ \midrule
			Sample & Metric    &  \multicolumn{1}{c}{MNP-BN} & \multicolumn{1}{c}{MNP-MPR} & \multicolumn{1}{c}{MNP-I} & \multicolumn{1}{c}{Naive} \\ \midrule
			in     & hit-rate  &                      0.495  &  0.473   & 0.000  &  0.000 \\
			in     & log-score &                      0.080  &  0.000   & 0.000  &  0.000 \\
			out    & hit-rate  &                      0.866  &  0.779   & 0.002  &  0.000 \\
			out    & log-score &                      0.026  &  0.000   & 0.000  &  0.000 \\ \toprule\toprule
			                               \multicolumn{6}{c}{6 margarine categories}                                                              \\ \midrule
			Sample & Metric    &  \multicolumn{1}{c}{MNP-BN} & \multicolumn{1}{c}{MNP-MPR} & \multicolumn{1}{c}{MNP-I} & \multicolumn{1}{c}{Naive} \\ \midrule
			in     & hit-rate  &                     1.000  &  0.944  &  0.725  &  0.259 \\
			in     & log-score &                     0.188  &  0.638  &  0.602  &  0.000 \\
			out    & hit-rate  &                     1.000  &  1.000  &  0.884  &  0.565 \\
			out    & log-score &                     0.009  &  0.721  &  0.891  &  0.370 \\ \midrule\midrule
			                               \multicolumn{6}{c}{50 margarine categories}                                                             \\ \midrule
			Sample & Metric    &  \multicolumn{1}{c}{MNP-BN} & \multicolumn{1}{c}{MNP-MPR} & \multicolumn{1}{c}{MNP-I} & \multicolumn{1}{c}{Naive} \\ \midrule
			in     & hit-rate  &                    0.058  &  0.001  &  0.000 &   0.000 \\
			in     & log-score &                    0.000  &  0.000  &  0.000 &   0.000 \\
			out    & hit-rate  &                    0.459  &  0.204  &  0.000 &   0.000 \\
			out    & log-score &                    0.000  &  0.000  &  0.000 &   0.000 \\ \bottomrule\bottomrule
		\end{tabular}%
		\begin{tablenotes}
			\footnotesize
			\item This table shows the p-values for the tests on the difference of the hit-rates and the difference of the log-scores between the MNP-FS and the benchmarks, for the detergent and margarine data sets discussed in Section~ \ref{sec: application}. See Table 3 for details.
		\end{tablenotes}
	\end{threeparttable}
	\label{tab:forecastmar_application_pvals}
\end{table}

\begin{table}[h!]
  \centering \small
  \caption{Hit-rate and log-score for empirical applications}
  \begin{threeparttable}
    \begin{tabular}{llrrrrr}
    \toprule \toprule
     \multicolumn{7}{c}{6 laundry detergent categories}\\
          \midrule
         Sample & Metric    & \multicolumn{1}{c}{ MNP-FS} & \multicolumn{1}{c}{MNP-BN} & \multicolumn{1}{c}{MNP-MPR} & \multicolumn{1}{c}{MNP-I} & \multicolumn{1}{c}{Naive} \\
          \midrule
    in    & hit-rate & 0.497 & 0.494\hspace{0.25cm} & 0.498\hspace{0.25cm} & 0.504\hspace{0.25cm} & 0.272$^{+}$ \\
    in    & log-score & -1.329 & -1.325$^{-}$ & -1.326$^{-}$ & -1.349$^{+}$ & -1.641$^{+}$ \\
    out   & hit-rate & 0.488 & 0.482\hspace{0.25cm} & 0.486\hspace{0.25cm} & 0.484\hspace{0.25cm} & 0.262$^{+}$ \\
    out   & log-score & -1.406 & -1.401$^{-}$ & -1.402\hspace{0.25cm} & -1.372\hspace{0.25cm} & -1.637$^{+}$ \\
              \midrule \midrule
          \multicolumn{7}{c}{50 laundry detergent categories}\\
          \midrule
           Sample & Metric    & \multicolumn{1}{c}{ MNP-FS} & \multicolumn{1}{c}{MNP-BN} & \multicolumn{1}{c}{MNP-MPR} & \multicolumn{1}{c}{MNP-I} & \multicolumn{1}{c}{Naive} \\
           \midrule
    in    & hit-rate & 0.217 & 0.226 & 0.213\hspace{0.25cm} & 0.165$^{+}$ & 0.058$^{+}$ \\
    in    & log-score & -3.308 & -3.306 & -3.409$^{+}$ & -3.725$^{+}$ & -3.792$^{+}$ \\
    out   & hit-rate & 0.208 & 0.206 & 0.204\hspace{0.25cm} & 0.154$^{+}$ & 0.055$^{+}$ \\
    out   & log-score & -3.350 & -3.350 & -3.456$^{+}$ & -3.752$^{+}$ & -3.761$^{+}$ \\

    \toprule \toprule
          \multicolumn{7}{c}{6 margarine categories}\\
          \midrule
           Sample & Metric    & \multicolumn{1}{c}{ MNP-FS} & \multicolumn{1}{c}{MNP-BN} & \multicolumn{1}{c}{MNP-MPR} & \multicolumn{1}{c}{MNP-I} & \multicolumn{1}{c}{Naive} \\
          \midrule
    in    & hit-rate & 0.468 & 0.466 & 0.468 & 0.478 & 0.426\hspace{0.25cm} \\
    in    & log-score & -1.451 & -1.448 & -1.452 & -1.453 & -1.581$^{+}$ \\
    out   & hit-rate & 0.624 & 0.624 & 0.624 & 0.634 & 0.584\hspace{0.25cm} \\
    out   & log-score & -1.269 & -1.264 & -1.270 & -1.270 & -1.323\hspace{0.25cm} \\
              \midrule \midrule
          \multicolumn{7}{c}{50 margarine categories}\\
          \midrule
           Sample & Metric    & \multicolumn{1}{c}{ MNP-FS} & \multicolumn{1}{c}{MNP-BN} & \multicolumn{1}{c}{MNP-MPR} & \multicolumn{1}{c}{MNP-I} & \multicolumn{1}{c}{Naive} \\
           \midrule
    in    & hit-rate & 0.360 & 0.345$^{+}$ & 0.335$^{+}$ & 0.232$^{+}$ & 0.106$^{+}$ \\
    in    & log-score & -2.813 & -2.826$^{+}$ & -2.894$^{+}$ & -3.462$^{+}$ & -3.547$^{+}$ \\
    out   & hit-rate & 0.344 & 0.331\hspace{0.25cm} & 0.323\hspace{0.25cm} & 0.209$^{+}$ & 0.095$^{+}$ \\
    out   & log-score & -2.844 & -2.863$^{+}$ & -2.939$^{+}$ & -3.505$^{+}$ & -3.561$^{+}$ \\
    \bottomrule \bottomrule
    \end{tabular}%
\begin{tablenotes}
\footnotesize
\item The MNP-FS is estimated with $\theta=(1.525,1,5)^\top$, which implies an equicorrelated prior mean for the covariance matrix. The results of the other models are identical to the results in Tables~\ref{tab:forecast} and \ref{tab:forecastmar}.
\end{tablenotes}
\end{threeparttable}
  \label{tab:forecastmar_application}
\end{table}
\end{document}